\documentclass[twocolumn,trackchanges]{aastex701}

\newcommand{\Mste}{$M_{\star}$}
\newcommand{\SSFR}{$\Sigma_{\rm SFR}$}
\newcommand{\sSSFR}{$\Sigma_{\rm sSFR}$}

\newcommand{\Menc}{$M_{\rm{enc}}$}
\newcommand{\SFRenc}{SFR$_{\rm{enc}}$}
\newcommand{\sSFRenc}{sSFR$_{\rm{enc}}$}
\newcommand{\SSFRenc}{$\Sigma_{\rm{SFR, enc}}$}
\newcommand{\sSSFRenc}{$\Sigma_{\rm{sSFR, enc}}$}

\newcommand{\Vmax}{$V_{\rm{max}}$}
\newcommand{\Vcir}{$v_{\rm{circ}}$}
\newcommand{\Vesc}{$v_{\rm{esc}}$}

\newcommand{\RE}{$R_{\rm{E}}$}
\newcommand{\ROUT}{$R_{\rm{out}}$}
\newcommand{\RENC}{$R_{\rm{enc}}$}

\newcommand{\neout}{$n_{\rm{e,\, out}}$}

\newcommand{\HB}{$\rm{H}\beta$}
\newcommand{\OIII}{[\ion{O}{3}]}
\newcommand{\HA}{$\rm{H}\alpha$}
\newcommand{\NII}{[\ion{N}{2}]}
\newcommand{\SII}{[\ion{S}{2}]}

\newcommand{\Mdot}{$\dot{M}_{\rm{out}}$}
\newcommand{\Mload}{$\eta_{m}$}
\newcommand{\Mdotfixed}{$\dot{M}_{\rm{out,\, fixed}}$}
\newcommand{\Mloadfixed}{$\eta_{m,\, \text{fixed}}$}

\usepackage{amsmath}
\usepackage{enumitem}
\usepackage{graphicx}

\begin{document}

\title{Electron Density of Ionized Gas Outflows: Insights from the MaNGA Survey}

\author[0000-0002-1272-9064]{Andrew Weldon}
\affiliation{Department of Physics and Astronomy, Texas A\&M University, 4242 TAMU, College Station, TX 77843-4242, USA}
\affiliation{George P. and Cynthia Woods Mitchell Institute for Fundamental Physics and Astronomy, Texas A\&M University, 4242 TAMU, College Station, TX 77843-4242, USA}
\email[show]{weldon@tamu.edu}  
\correspondingauthor{Andrew Weldon}	
\author[0000-0001-7503-8482]{Casey Papovich}
\affiliation{Department of Physics and Astronomy, Texas A\&M University, 4242 TAMU, College Station, TX 77843-4242, USA}
\affiliation{George P. and Cynthia Woods Mitchell Institute for Fundamental Physics and Astronomy, Texas A\&M University, 4242 TAMU, College Station, TX 77843-4242, USA}
\email{papovich@tamu.edu} 
\author[0000-0003-3256-5615]{Justin Spilker}
\affiliation{Department of Physics and Astronomy, Texas A\&M University, 4242 TAMU, College Station, TX 77843-4242, USA}
\affiliation{George P. and Cynthia Woods Mitchell Institute for Fundamental Physics and Astronomy, Texas A\&M University, 4242 TAMU, College Station, TX 77843-4242, USA}
\email{jspilker@tamu.edu} 
\author[0000-0001-5448-1821]{Robert C. Kennicutt, Jr.}
\affiliation{Department of Physics and Astronomy, Texas A\&M University, 4242 TAMU, College Station, TX 77843-4242, USA}
\affiliation{George P. and Cynthia Woods Mitchell Institute for Fundamental Physics and Astronomy, Texas A\&M University, 4242 TAMU, College Station, TX 77843-4242, USA}
\affiliation{Department of Astronomy and Steward Observatory, University of Arizona, 933 N. Cherry Avenue, Tucson, AZ 85721, USA}
\email{rck@arizona.edu} 

\begin{abstract}

We investigate the properties of ionized gas outflows in nearby star-forming galaxies from the final Data Release of the Mapping Nearby Galaxies at Apache Point Observatory (MaNGA) Survey. Using spatially resolved spectroscopy, we search for signatures of ionized outflows within physically motivated outflow apertures. We find significant evidence for additional broad \HA, \NII$\lambda\lambda$6550,6584, and [\ion{S}{2}$]\lambda\lambda$6718,6733 emission-line components in 115 galaxies, $\sim$3\% of star-forming MaNGA galaxies. Our analysis suggests that outflow electron densities (\neout), mass-loss rates, and mass-loading factors are related to both the properties of the regions that launch and global galactic properties. In particular, we find correlations between \neout\ and both the enclosed ($-$3.4$\sigma$) and global ($-$2.4$\sigma$) star-formation-rate surface densities, such that lower surface density regions tend to drive denser ionized gas outflows. We find several $>$$2\sigma$ correlations between the mass-loading factor and galactic properties using individual \neout\ estimates, while more significant correlations ($>$$3\sigma$) only emerge when \neout\ is fixed to the median value of the sample. Combining with literature \neout\ measurements out to $z \sim 1.9$, \neout\ exhibits no significant redshift evolution, in contrast to the strong evolution observed for gas electron densities of the interstellar medium (ISM). Ionized outflows remain denser than the ISM by a factor of $\sim$8.5 at $z \sim 0$ and $\sim$1.6 at $z \sim 1.9$. These results provide new constraints for feedback models and highlight the need for high-resolution IFU observations and synthetic line diagnostics from simulations to investigate the connection between outflow properties and local ISM conditions.

\end{abstract}



\section{Introduction} 
\label{sec:intro}

The interplay between gas accretion, star formation, and feedback governs the evolutionary path of galaxies. The accretion of gas from the intergalactic medium (IGM) serves to fuel the formation of new stars and the feeding of central supermassive black holes. At the same time, feedback regulates these processes by injecting mass, energy, and momentum back into the interstellar medium (ISM). These feedback processes, driven by stellar winds, supernova explosions, and active galactic nuclei (AGN), can mix, heat up, or drive gas outflows, and are essential for reproducing a wide range of key observables in galaxy evolution. Through these mechanisms, feedback contributes to the inefficiency of star formation in both low-mass galaxies, due to effective outflows, and high-mass galaxies, by AGN feedback that heats or removes circumgalactic gas. In modern cosmological simulations, feedback from stellar winds, supernovae, and AGN is an essential ingredient to suppress star formation and prevent the overproduction of stellar mass \citep[e.g.,][]{Hopkins18, Pillepich18}.

Observationally, the impact of feedback is reflected in several fundamental galaxy scaling relations. The small intrinsic scatter of the star-forming main sequence implies that star formation is regulated by processes that act across diverse galaxy populations \citep[e.g.,][]{Noeske07, Speagle14}, consistent with self-regulation by feedback-driven outflows. These outflows likely play a similar role in establishing the mass-metallicity relation by regulating the retention and recycling of enriched material in galaxies \citep[e.g.,][]{Tremonti04, Dalcanton07, Finlator08} and in enriching the circumgalactic medium (CGM) and IGM with metals \citep[e.g.,][]{Oppenheimer06, Scannapieco06, Shen10}. Outflows also appear to be an important factor in the formation of low-column-density channels in the ISM, facilitating the escape of ionizing photons \citep[e.g.,][]{Gnedin08, Ma16, Reddy22}. Given the variety of consequences that feedback has on galaxy evolution, characterizing their physical properties is essential for understanding galaxy evolution.

Over the past decade, numerous observational studies have investigated the properties of galactic-scale outflows across a broad range of redshifts and gas phases. Multiwavelength observations have shown that outflows have a multiphase structure. Outflows are detected in X-ray emitting hot ($\sim$10$^{6-7}$ K) gas, rest-optical emission and rest-UV absorption lines that trace warm ionized ($\sim$10$^{4}$ K) and cool neutral ($\sim$10$^{3}$ K) gas, and molecular CO and OH emission and absorption from cold ($\lesssim$100 K) gas \citep[see reviews by][]{Heckman17, Rupke18, Veilleux20}. At low redshifts ($z \lesssim$ 0.25), studies of galactic-scale outflows driven by stellar feedback have found that outflow velocity increases with the stellar mass, star-formation rate (SFR), and star-formation-rate surface density (\SSFR) of their host in both the warm ionized \citep[e.g.,][]{Arribas14, Cicone16, Avery21} and cool neutral phases \citep[e.g.,][]{Chen10, Chisholm15, Heckman15}. This suggests a tight connection between feedback strength and star-forming activity. Galactic-scale outflows are expected to be more common and effective at removing material at redshifts $z \sim 1-3$, during the peak of the cosmic star formation history, when feedback from star formation was at its strongest \citep{Madau14}. Indeed, both the detection rate and velocity of outflows appear enhanced in $z \sim 2$ galaxies compared to nearby galaxies \citep[e.g.,][]{Sugahara17}. However, the existence and strength of correlations between outflow velocity and star formation properties at these redshifts remain debated, with some studies reporting significant trends \citep[e.g.,][]{Kornei12, Bordoloi14, Davies19}, while others find no clear dependence \citep[e.g.,][]{Law12, Prusinski21, Weldon24}.

Despite several decades of observational efforts, many fundamental properties of outflows beyond their kinematics remain poorly understood. A key property of outflows is the amount of mass ejected (\Mdot) relative to the galaxy’s star-formation rate, known as the  mass-loading factor (\Mload = \Mdot/SFR). In star-forming galaxies, \Mload\ is thought of as a proxy for outflow efficiency and provides constraints on models of feedback regulation and chemical enrichment \citep[e.g.,][]{Dave11, Muratov15, Nelson19}. However, estimating \Mload\ remains observationally challenging due to its dependence on multiple poorly constrained outflow properties. For example, outflows traced by absorption lines require assumptions of the outflow geometry, the column density of the outflowing gas, and the absorption contribution from the ISM and faint satellite galaxies, which have limited the number of absorption-line studies that provide robust constraints on \Mload\ \citep[][]{Chisholm18, Xu23a}. On the other hand, outflows traced by emission lines depend on similar unconstrained properties (e.g., outflow geometry), but also offer the advantage of direct constraints on some properties, such as the electron density of the outflowing gas (\neout). Previous studies have investigated \Mload\ of warm ionized gas outflows, finding correlations with stellar mass and the star-formation properties of the host galaxy \citep[e.g.,][]{Arribas14, Avery21, Marasco23}, in broad agreement with theoretical predictions.

While warm ionized outflow studies have focused on constraining \Mload\ and its scaling with galactic properties, little exploration has been given to how the electron density and spatial extent of the outflowing gas may vary with the properties of their host. In particular, the electron density of the ISM is reported to increase with \SSFR\ \citep[e.g.,][]{Shimakawa15, Reddy23, Li25}, suggesting that \neout\ may similarly correlate with host galaxy properties. Recently, \cite{Xu23a} reported a correlation between \SSFR\ and the electron density of \textit{cool, neutral} outflowing gas measured using rest-frame UV lines in 45 local star-forming galaxies ($z < 0.18$), suggesting that a similar trend may also occur for ionized gas outflows. A persistent observational limitation in measuring the electron density of ionized gas outflows is detecting and resolving the faint, broad emission components of density-sensitive lines, such as the [\ion{S}{2}$]\lambda\lambda$6718,6733 doublet. Previous studies often adopt a single, representative \neout\ for entire samples of galaxies that span a broad range in stellar mass, SFR, and \SSFR\ \citep[e.g.,][]{Arribas14, Avery21, Xu22, Weldon24}. As \Mload\ scales inversely with \neout, these assumptions can introduce substantial uncertainties into \Mload\ and potentially affect the observed physical trends.

The emergence of large integral field unit (IFU) surveys of local galaxies now offers a powerful opportunity to systematically investigate ionized outflows. Surveys such as MaNGA \citep{Bundy15}, SAMI \citep{Croom12}, and CALIFA \citep{Sanchez12} provide spatially resolved spectroscopy for thousands of galaxies across a wide range of stellar masses, morphologies, and SFRs. With full spectral coverage of rest-optical emission lines, moderate spatial resolution, and high sensitivity, these surveys enable the detection and characterization of spatially extended outflows. By combining statistical power with detailed spectral measurements, IFU surveys offer a unique opportunity to study the physical properties of outflows and directly test the predictions of feedback models in the nearby Universe.

In this paper, we present an analysis of ionized gas outflows in star-forming galaxies using the final data release of the Mapping Nearby Galaxies at Apache Point Observatory (MaNGA) survey. Leveraging the IFU observations, we estimate outflow velocity, electron density, and extent on an individual galaxy basis. Our goals for this study are to explore which, if any, of these outflow properties correlate with the properties of their host. The outline of this paper is as follows. In Section \ref{sec:Sample}, we introduce the MaNGA survey, describe the sample selection, and outline the measurements of galaxy properties. Section \ref{sec:methods} describes our methodology for fitting galaxy spectra to characterize the presence of outflows. The main results on the correlations between outflow and galaxy properties are presented in Section \ref{sec:results} and further discussed in Section \ref{sec:discussion}. Finally, we present the summary and main conclusions of this work in Section \ref{sec:con}. Throughout this paper, we adopt a \cite{Chabrier03} initial mass function (IMF) and a standard cosmology with $\Omega_{\Lambda}$ = 0.7, $\Omega_{M}$ = 0.3, and $\rm{H}_{0}$ = 70 km s$^{-1}\ \rm{Mpc}^{-1}$. All wavelengths are presented in the vacuum frame. 

\section{Sample} 
\label{sec:Sample}

\subsection{MaNGA Survey}

The work presented in this paper utilizes the final data release of the SDSS--MaNGA survey \citep[DR17;][]{Abdurrouf22}, providing spatially resolved spectroscopy for over 10,000 galaxies in the nearby Universe (0.01 $<$ $z$ $<$ 0.15). All galaxies were observed with optical fibers \citep{Drory15} fed to the Baryon Oscillation Spectroscopic Survey spectrographs \citep{Smee13} on the 2.5m Sloan Telescope at Apache Point Observatory \citep{Gunn06}. The MaNGA survey used bundles of between 19 and 127 fibers to obtain spectra across a 2D, hexagonal field of view. Each spectrum covers a wavelength range between $\sim$3600--10400~\r{A} with a spectral resolution that ranges from R $\sim$ 1400 (214 km/s) at 4000~\r{A} to R $\sim$ 2600 (115 km/s) at 9000~\r{A}.

MaNGA targeted galaxies to create an unbiased sample of the galaxy population in the local Universe. To achieve this, the survey is composed of three subsamples: (i) a primary sample with radial coverage out to 1.5 effective raii (\RE); (ii) a color-enhanced supplement to the primary sample adding green valley, low-mass red, and high-mass blue galaxies; and (iii) a secondary sample with radial coverage out to 2.5\RE. Further details on the observing strategy, survey design, and sample selection are given in \cite{Law15}, \cite{Yan16}, \cite{Wake17}.

For our analysis, we use the reduced MaNGA data cubes with linear wavelength sampling from the Data Reduction Pipeline \citep[DRP;][]{Law16} and 2D maps of stellar kinematics and emission-line properties from the Data Analysis Pipeline \citep[DAP;][]{Belfiore19, Westfall19}. Specifically, we use results from the hybrid binning model (HYB10-MILESHC-MASTARSSP), where the stellar kinematics are performed on Voronoi-binned spectra using the MILES stellar library, and emission lines are fitted for each spaxel using the MaStar SSP library \citep{Maraston20}. In addition, our analysis benefits from the Pipe3D Value-Added Catalog, which provides spatially resolved 2D maps of galactic properties and emission-line fluxes, and a rich set of integrated measurements for MaNGA galaxies\footnote{\url{https://www.sdss4.org/dr17/manga/manga-data/manga-pipe3d-value-added-catalog/}} \citep{Sanchez22}.

\subsection{Sample Selection}
\label{subsec:sample}

In order to maximize the potential outflow sample, we select star-forming galaxies from the MaNGA survey using the following minimal criteria. We begin with the 10,782 objects analyzed by DAP, provided in the DAPall catalog, removing 1813 objects with an assigned DAP quality control flag or flagged as a ``Close Pair and Merger'' target\footnote{\url{https://www.sdss4.org/dr17/manga/manga-target-selection/ancillary-targets/close-pairs-and-mergers/}}. Next, we cross-match and remove 339 AGNs present in the MaNGA-AGN Value-Added Catalog \citep{Comerford24}, which identifies AGN based on Wide-field Infrared Survey Explorer mid-infrared colors, ultrahard X-ray detections from the Swift observatory’s Burst Alert Telescope all-sky survey, radio detections in the NRAO Very Large Array Sky Survey \citep[NVSS;][]{Condon98} and the Faint Images of the Radio Sky at Twenty centimeters survey \citep[FIRST;][]{Becker95}, or broad \HA\ emission lines in SDSS spectra. Next, we cross-match the galaxies to the Pipe3D catalog, removing 245 galaxies with an assigned Pipe3D quality control flag. Finally, we use Pipe3D emission-line ratios measured within a 2.5 arcsec aperture at the galaxy center to classify galaxies on the BPT-NII diagnostic diagram \citep{BPT}. We classify and remove 4830 galaxies that lie above the \cite{Kauffmann03} separation curve. These criteria result in a parent sample of 3555 star-forming galaxies.

\section{Search for Ionized Outflows} 
\label{sec:methods}

In this section, we describe our procedure for identifying ionized gas outflows. 
To estimate \neout, we search for signatures of ionized outflowing gas in the \HA, \NII$\lambda\lambda$6550,6584, and [\ion{S}{2}$]\lambda\lambda$6718,6733 emission lines. We exclude the \HB\ and \OIII$\lambda\lambda$4960,5008 emission lines to minimize systematic uncertainties from their large wavelength separation from \HA/\NII/\SII, their greater sensitivity to dust attenuation and flux-calibration uncertainties, and the possibility that \OIII\ traces a higher ionization gas component with different kinematics \citep[e.g.,][]{Yu22}. The extraction of 1D spectra from the data cubes and stellar continuum subtraction are described in \ref{subsec:reduce}. A preliminary search for outflows from the parent sample is discussed in \ref{subsec:lmfit}. In \ref{subsec:cog}, we discuss how the radial extent of outflow candidates is measured. Finally, we confirm the presence of ionized outflows within the outflow radius in \ref{subsec:mcmc}.

\subsection{1D Spectra and Continuum Subtraction}
\label{subsec:reduce}

We utilize the high sensitivity and spatially resolved spectroscopy of the MaNGA survey to search for evidence of ionized gas outflows from galaxies. In particular, we extract 1D spectra from elliptical apertures using the Petrosian r-band ellipticities and position angles from the NASA-Sloan Atlas (NSA) catalog. For each spaxel within a given aperture, their science and error spectra are corrected for the large-scale velocity gradient of the galaxy using the \HA\ velocity field map provided by the DAP and interpolated onto a common grid with a wavelength spacing of $\Delta\lambda$ = 1~\r{A}. Next, any spaxel in the velocity map flagged as ``DONOTUSE'' or ``UNRELIABLE'' by the DRP or DAP is removed. Integrated spectra are then produced by summing over the remaining spaxels, while the errors are added in quadrature and corrected for the covariance between spatially adjacent spaxels following Equation (9) of \cite{Law16}.

In order to search for ionized outflows, we first subtracted the stellar continuum from the spectra, because the continuum can mask faint, broad wings in line profiles that may be associated with an outflow component. The stellar continuum subtraction is performed through multi-component spectral modeling using the \texttt{pPXF} software \citep{Cappellari17, Cappellari23}. The models consist of: (1) a stellar continuum of E-MILES \citep{Vazdekis16} stellar population model templates with four Gauss-Hermite moments, (2) emission lines fitted as single Gaussians with velocity offsets and widths tied between lines, and (3) an additive third-order polynomial to account for the possible mismatch between the spectral templates and the spectra that change slowly with wavelength. For the emission lines, the Balmer lines are fit as a single component with ratios tied to their intrinsic values, gas reddening is left as a free parameter, and the [\ion{O}{2}] and \SII\ doublets are limited to their physically allowed range. Using the best-fit \texttt{pPXF} model, we subtract the stellar continuum from the galaxies’ spectra to produce emission-line-only spectra.

Detecting broad components of rest-optical emission lines is sensitive to the residual continuum flux around the lines. Any poorly subtracted stellar continuum from the best-fit \texttt{pPXF} model may serve to artificially enhance or suppress these broad features. To improve the continuum subtraction around our lines of interest, we adopt a similar procedure to that of \cite{Marasco23}. For each emission line, we take a 140~\AA\ region centered on the line. We then mask pixels that encompass any potential broad or narrow emission component and fit a third-order polynomial to the masked, emission-line-only spectrum in this window (masking also any other emission lines in the window).  This fitted ``local'' continuum is then subtracted, leaving us with a continuum-subtracted, emission-line-only spectrum in a wavelength window appropriate for each line. We perform this additional continuum correction for windows centered on the \HB\ emission line, the \HA/\NII\ complex, and the \SII\ doublet.

\begin{figure*}
  \includegraphics[width=\linewidth, keepaspectratio]{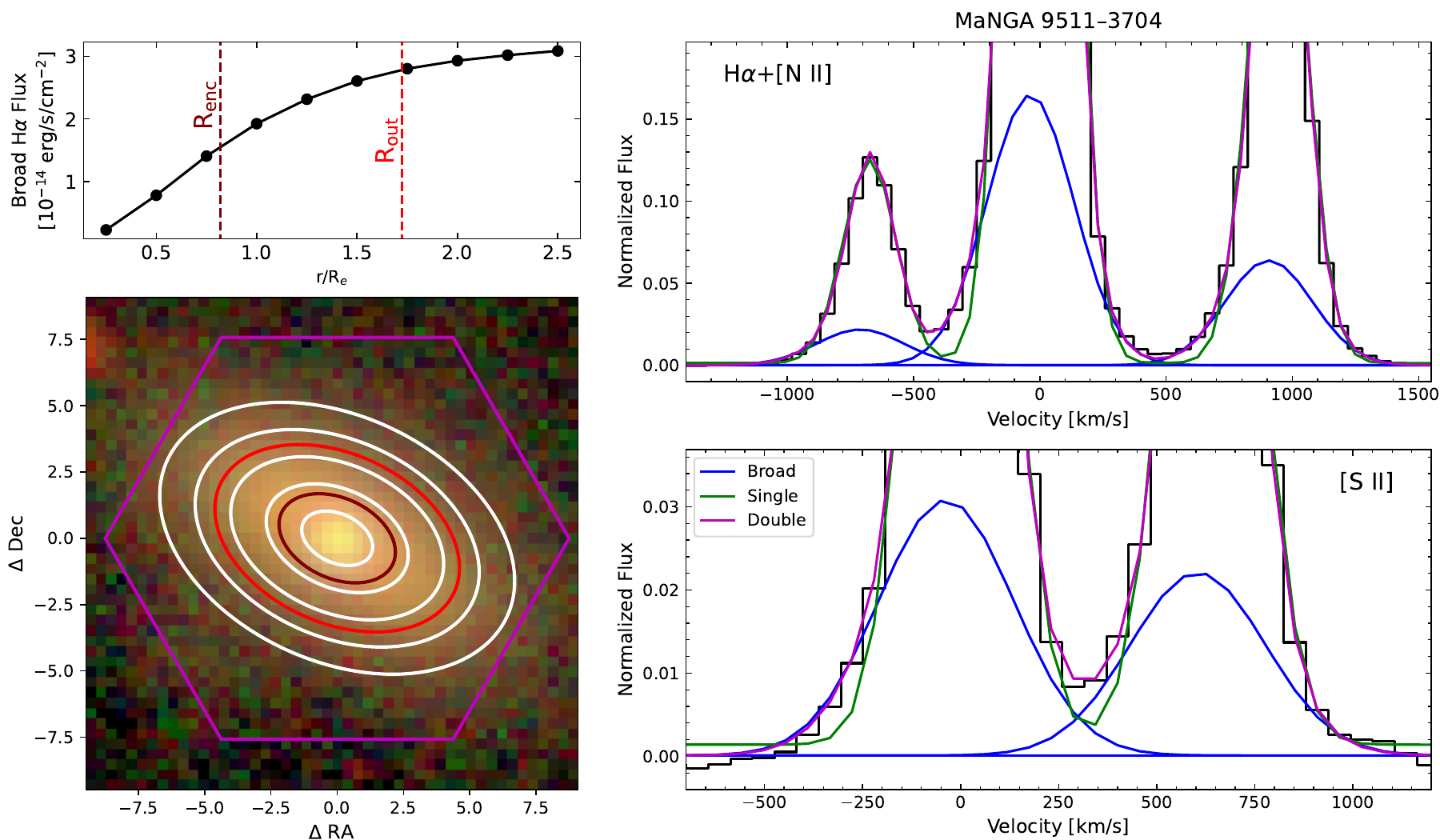}
  \vspace{-0.6cm}
  \caption{Overview of the search for ionized gas outflows. Top left: broad-component \HA\ flux in apertures of semi-major axes ranging from 0.25\RE\ up to 2.5\RE\, in 0.25\RE\ increments. The red (maroon) vertical line marks the radius enclosing 90\% (50\%) of the broad-component \HA\ flux. Bottom left: a true-color image created using the SDSS g-, r- and i-band images. The hexagonal MaNGA footprint is shown in magenta, the white ellipses show apertures from 0.5\RE\ to 2.5\RE\ in 0.5\RE\ increments, and the red (maroon) ellipse shows the ``outflow'' (``enclosed'') aperture. Right panels: Example double-Gaussian fit jointly to \HA, \NII$\lambda\lambda$6550,6584 and [\ion{S}{2}$]\lambda\lambda$6718,6733 within the outflow aperture. The green curve shows the single Gaussian fit. The purple curve shows the double Gaussian fit that includes a narrow and broad (blue curve) component.}
  \label{fig:example}
\end{figure*}

\subsection{Preliminary Search}
\label{subsec:lmfit}

The spectra of galaxies hosting ionized gas outflows are characterized by an underlying, broad emission component superimposed onto the systemic narrow emission from the host. In such cases, strong rest-optical emission lines can be decomposed into narrow kinematic components tracing virial motions within the host and broad kinematic components associated with the outflowing gas. However, decomposing emission lines becomes significantly more difficult when the outflow is faint relative to the emission from the host, when the bulk of the outflowing gas is near the systemic velocity of the host, or when the outflow does not align well with our line-of-sight.

With these considerations in mind, we perform a preliminary search to identify galaxies possibly hosting detectable ionized gas outflows. As the extent of the outflowing gas likely varies between galaxies due to the location of the driving mechanism and the time elapsed since the outflow launched, we begin by considering several aperture sizes. Specifically, for each galaxy, we extract integrated spectra from elliptical apertures with semi-major axes of 0.5\RE, 1.0\RE, 1.5\RE, and 2.0\RE\ following the procedure in Section \ref{subsec:reduce}.

We perform two separate fits to the line profiles of \HA, \NII$\lambda\lambda$6550,6584, and [\ion{S}{2}$]\lambda\lambda$6718,6733 using \texttt{lmfit} \citep{lmfit}, a nonlinear least-squares fitting Python package. The first fit models the emission lines as single Gaussians with a constant background, while the second fit models the lines as a set of two Gaussian components with a constant background. In both models, each Gaussian is fit with an amplitude and ``intrinsic'' FWHM, defined as:
\begin{equation}
    \text{FWHM}_{\text{obs}}^{2} = \text{FWHM}_{\text{int}}^{2} + \text{FWHM}_{x}^{2}(\lambda)
\end{equation}
where FWHM$_{\text{obs}}$ is the observed FWHM of the emission lines and FWHM$_{x}$ is the $\lambda$-dependent instrumental broadening. To simplify the fits and reduce degeneracies, we apply the following assumptions: (1) the single (narrow) components share a common FWHM and redshift, (2) the broad components share a common FWHM and velocity offset, and (3) the [\ion{N}{2}]$\lambda$6585/[\ion{N}{2}]$\lambda$6550 flux ratio is fixed to 2.93, consistent with atomic physics \citep{Osterbrock89}. Additionally, during the fitting, we constrain the amplitude of each line component to be positive and restrict the broad FWHM to be at least 1.2$\times$ larger than the narrow FWHM. 

This preliminary search aims to further refine our sample selection, providing an efficient alternative to the more intensive search described in Section \ref{subsec:cog}. As such, we adopt more relaxed criteria than typically seen in the literature to assess whether a broad component is detected. Specifically, we evaluate the improvement of the double Gaussian model over the single Gaussian model using the Bayesian Information Criterion \citep[BIC;][]{Schwarz78}. The BIC is defined as: 
\begin{equation}
    \text{BIC} = \chi^{2} + k \text{ln}\left(n \right),
\end{equation}
where $\chi^{2}$ is the chi squared of the fit, $k$ is the number of parameters used in the fit, and $n$ is the number of points used in the fit. We consider evidence for the presence of a broad component if the fit shows an improvement given by the difference in the BICs, $\Delta$BIC = BIC$_{\rm{single}}$ $-$ BIC$_{\rm{double}}$ $>$ 0, and if the flux ratio of the broad-to-narrow components (BNR) is at least 1\% in each of the four emission lines. Using these thresholds, 863 galaxies exhibit a possible broad component in at least one of the elliptical apertures.

\subsection{Outflow Extent}
\label{subsec:cog}

Studies of ionized gas outflows have primarily relied on a single integrated spectrum per galaxy, offering little insight into the spatial extent of these outflows. The spatially resolved spectroscopy provided by the MaNGA survey allows us to explore outflows across a galaxy. In this section, we take advantage of this capability to estimate the spatial extent of outflows by constructing a radial curve of growth of the broad \HA\ component flux; see Figure \ref{fig:example}.

To trace the spatial extent of ionized outflows in each galaxy, we extract integrated spectra from a series of elliptical apertures with semi-major axes ranging from 0.25\RE\ up to 2.5\RE\, in 0.25\RE\ increments, following the procedure outlined in Section \ref{subsec:reduce}. Next, we perform an initial fit to the emission-line profiles of \HA, \NII$\lambda\lambda$6550,6584 and [\ion{S}{2}$]\lambda\lambda$6718,6733 with the single- and double-Gaussian models as described in Section \ref{subsec:lmfit}. The best-fit values from these fits are then used as initial values for the final fit using \texttt{emcee} \citep{emcee}, a Python-based Markov Chain Monte Carlo (MCMC) ensemble sampler. To ensure that the broad components trace a kinematically distinct component rather than artifacts of the fitting process, we apply the following physically motivated priors: all line component amplitudes must be positive, the broad FWHM must be at least 1.2$\times$ larger than the narrow FWHM, and the velocity centroids of the narrow and broad components are restricted to within $\pm$100 km s$^{-1}$ of their initial values, as found by similar studies \citep[e.g.,][]{Wood15}. Additionally, the ratio of [\ion{S}{2}$]\lambda\lambda$6718,6733 is limited to the physically allowed range [0.45, 1.45] (see Section \ref{subsubsec:ne_scaling}). From the resulting posterior distributions, we adopt the median values as the best-fit parameters and estimate uncertainties using the 16th and 84th percentiles.

To determine whether a broad component is detected, we apply the following criteria:
\begin{enumerate}
    \item The need for a second emission component. We select cases with $\Delta$BIC$\,>\,$100, indicating that the double-Gaussian model is strongly preferred to the single-Gaussian model, and a broad FWHM significantly larger ($>3\sigma$) than the single FWHM. 
    \item Clear detection of two emission components. We select cases where the broad component of each emission line is reliably detected, with a S/N $>$ 3 in both its amplitude and integrated flux, and a broad-to-narrow flux ratio of at least 5\%.
\end{enumerate}

Our choice to adopt a $\Delta$BIC $>$100 is motivated by other work \citep{Chu22,Chu25} to ensure a more robust sample. Other results in the literature use lower values, e.g., $\Delta$BIC $>$ 10. However, the integrated MaNGA spectrum achieves very high signal-to-noise by coadding many spaxels, with an average peak S/N of $\gtrsim$ 100 for \HA, \NII, and \SII\ in the integrated spectra. While the high S/N improves our ability to detect broad emission components, it also increases the sensitivity to any fluctuations away from a Gaussian shape (e.g., imperfect continuum subtraction). As a result, the double-Gaussian model can reach a low BIC$_{\rm{double}}$ value (i.e., large $\Delta$BIC) through fitting such small fluctuations that do not correspond to physically meaningful broad emission. We find that a conservative threshold of $\Delta$BIC = 100 produces a reliable sample. This threshold prioritizes robustness, but may bias results by excluding low-S/N spectra or outflows with weak fluxes or low velocities. We note that our main results do not change if we adopt a $\Delta$BIC value of 50.

With these criteria, we construct the radial curve of growth of the broad-component \HA\ flux for each galaxy. Following \cite{Avery21}, we adopt the radius enclosing 90\% of the broad-component \HA\ flux as the outflow radius \ROUT. In total, we measure an outflow radius for 133 galaxies from the outflow candidate sample. 

\subsection{Ionized Outflows}
\label{subsec:mcmc}

With estimates of \ROUT\ in hand, we can assess whether each galaxy exhibits evidence of ionized gas outflows within this region. Following the procedure in Section \ref{subsec:reduce}, we extract integrated spectra from elliptical apertures with semi-major axes of \ROUT. We then fit the emission-line profiles with single- and double-Gaussian models using \texttt{emcee}, as outlined in Section \ref{subsec:cog}.

In addition to the outflow selection criteria described in Section \ref{subsec:cog}, we remove galaxies with their emission dominated by diffuse ionized gas (DIG). Warm ionized gas traced by rest-optical emission lines primarily originates from HII regions but also includes DIG located outside these regions, which may be kinematically offset from the large-scale motion of the galaxy \citep{Bizyaev17}. As a result, DIG may contribute to or be misidentified as the broad emission component. To avoid this potential contamination, we follow the approach of \cite{Avery21} and remove galaxies where DIG dominates the emission within \ROUT. Using the \HA\ equivalent width (EW) maps from the MaNGA DAP, galaxies are identified as DIG-dominated if more than half the spaxels within \ROUT\ have \HA\ EW $<$ 3 \citep[e.g.,][]{Lacerda18, Vale19}.

Of the 133 galaxies with a measured \ROUT, 18 no longer satisfy the outflow criteria and two are classified as DIG-dominated, leaving us with a final outflow sample of 115 galaxies\footnote{If the \SII\ lines are consider jointly, rather than separately, for the second outflow criteria, one additional galaxy would be included in the final sample.}.

\subsection{Galaxy Properties}

In this study, we explore how the properties of ionized gas outflows relate to several global galaxy properties (e.g., \Mste, SFR, \SSFR). We take integrated stellar masses (\Mste) and dust-corrected SFRs from the Pipe3D Value-Added Catalog, converting them to our adopted cosmology (h = 0.70), \cite{Chabrier03} IMF (converting from a \cite{Salpeter55} IMF), and \HA$-$SFR calibration. Using these values, we calculate the specific star formation rate (sSFR = SFR/\Mste), which acts as a tracer of both mechanical and gravitational potential energy. Along with SFR, the mechanisms that launch outflows may be enhanced in galaxies with compact star formation; thus, we estimate the star-formation-rate surface density defined as \SSFR\ = SFR/(2$\pi R_{\rm{E}}^{2}$). At a given \SSFR, outflows may be more effectively launched from a shallow galaxy potential (i.e., low stellar mass) relative to a deep potential \citep{Reddy22}. To examine the frequency of galaxies with observed outflows on both \SSFR\ and the galaxy potential, we define the specific star-formation-rate surface density (\sSSFR) as \sSSFR\ = SFR/(2$\pi R_{\rm{E}}^{2} M_{\star}$).

Finally, outflows are theorized to emerge from galaxies in a biconical structure \citep{Heckman90, Katz93}; thus, their properties may vary with inclination. Using the NSA semi-minor to semi-major axis ratio (b/a), we estimate inclination for 82 galaxies, with S\'{e}rsic indices $<2$, as:
\begin{equation}
    \text{sin}(i) = \sqrt{\frac{1- (b/a)^{2}}{1 - q_{0}^{2}}}
\end{equation}
where $q_{0}$ = 0.13 \citep{Giovanelli94} is the assumed intrinsic axial ratio.

In addition to global galaxy properties, studies have shown that outflow properties are often closely tied to the local regions that launch the outflowing gas \citep[e.g.,][]{Davies19, Roberts20, Chu22}. To investigate this possibility, we measure galactic properties ``enclosed'' within the radius enclosing 50\% of the broad-component \HA\ flux (\RENC). Leveraging the 2D maps of galactic properties and emission-line fluxes in the Pipe3D Value-Added Catalog, we calculate enclosed stellar masses (\Menc), and \HA\ and \HB\ fluxes by integrating over the stellar mass surface density and emission-line flux maps, respectively. Enclosed SFRs (\SFRenc) are estimated from dust-corrected \HA\ luminosities, using the \HA\ to \HB\ ratio, assuming an intrinsic ratio of 2.86 and the \cite{Cardelli89} extinction curve, with the calibration of \cite{Kennicutt12}, converted for a \cite{Chabrier03} IMF. Finally, enclosed sSFR, \SSFR, and \sSSFR\ are calculated following the same definitions as above, using \Menc, \SFRenc, and \RENC.

\begin{figure*}
  \includegraphics[width=\linewidth, keepaspectratio]{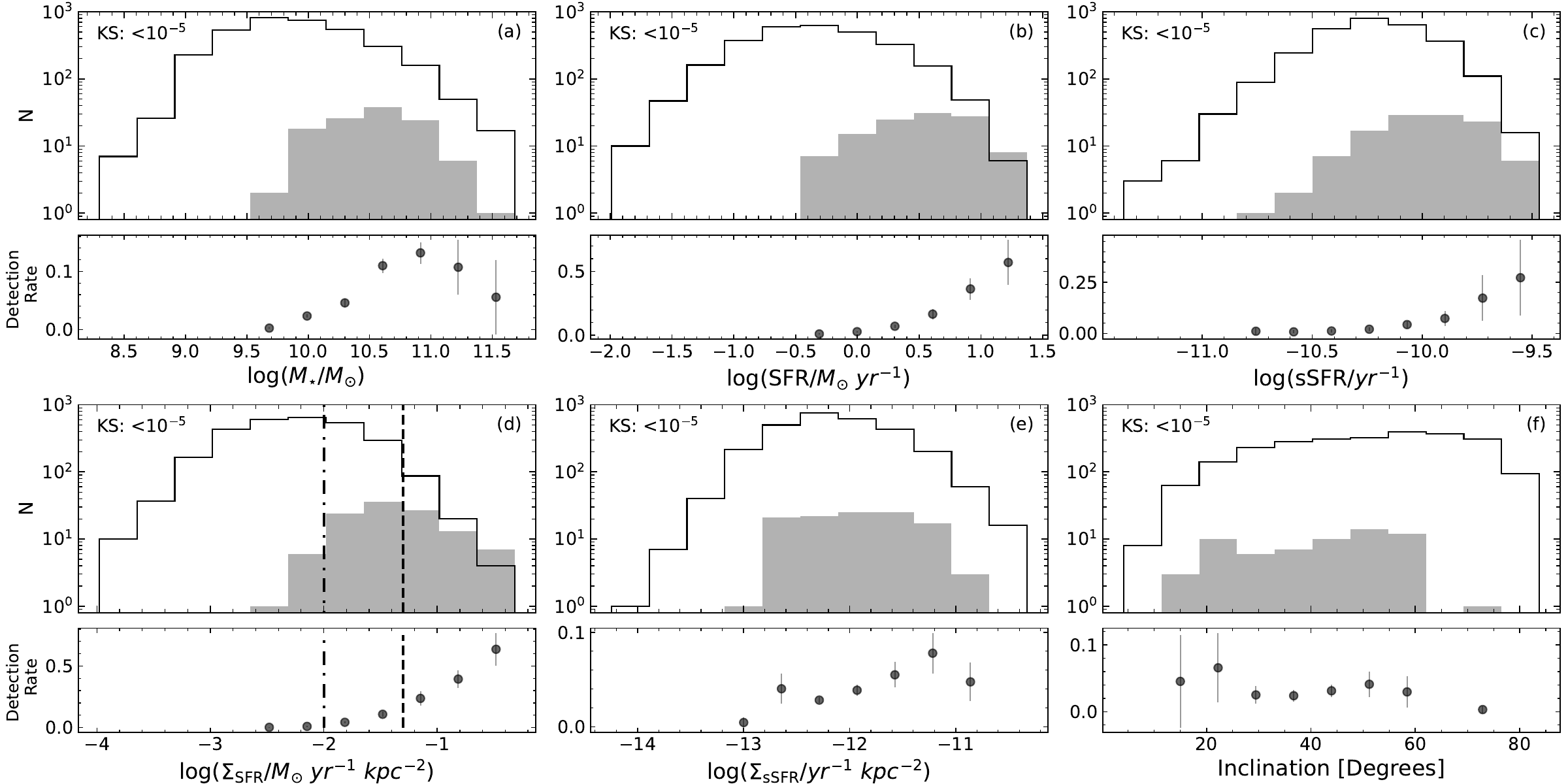}
  \vspace{-0.5cm}
  \caption{Top: Distribution of galactic properties for the 115 galaxies with detected broad, outflow components (solid gray) and the remaining galaxies (open histograms). The p-value of a KS test between galaxies with a broad component and the remaining galaxies is shown in the upper left corner. Bottom: Detection rate of ionized outflows as a function of host galactic properties. Panel (a): stellar mass, Panel (b): SFR, Panel (c) specific-SFR, Panel (d) SFR surface density, Panel (e) specific-SFR surface density, Panel (e) inclination. The vertical dotted-dashed in panel (d) marks the local SFR surface density threshold proposed by \cite{Roberts19} and \cite{Lopez19} for launching outflows, while the dashed line marks the global SFR surface density threshold proposed by \cite{Heckman02}.}
  \label{fig:incidence}
\end{figure*}

\section{Results} 
\label{sec:results}

\subsection{Incidence of Outflows}
\label{subsec:detection}

As outlined in Section \ref{subsec:sample}, we select 3,555 star-forming galaxies from the 10,782 galaxies observed as part of the MaNGA survey without imposing any cuts based on emission-line signal-to-noise. Of these, 863 galaxies exhibit potential for a second, broad emission component after decomposing the \HA, \NII$\lambda\lambda$6550,6584 and [\ion{S}{2}$]\lambda\lambda$6718,6733 lines using \texttt{lmfit}, a least-squares fitter. From this outflow candidate sample, significant evidence for broad emission is confirmed for 115 galaxies using \texttt{emcee}, an MCMC ensemble, within a physically motivated outflow aperture. Overall, broad emission is detected in 3\% (13\%) of the parent (outflow candidate) sample. The detection rate of the outflow candidate sample is similar to those reported for ionized outflow studies with earlier MaNGA data releases. 
For example, \cite{Pinto19} and \cite{Avery21} detected broad emission in 7\% and 12\% of their ``analyzed'' samples after applying cuts to their respective parent samples. 

Here, we explore the properties of galaxies hosting ionized outflows and compare them to those without detected outflows. The top panels of Figure \ref{fig:incidence} show distributions of \Mste, SFR, sSFR, \SSFR, \sSSFR, and inclination for the outflow sample (gray histograms) and the remaining parent sample (open histograms). There is a clear offset between the two samples, with galaxies hosting outflows shifted toward higher masses, higher star-formation activity, and lower inclinations (i.e., face-on). To test the significance of these apparent differences, we perform a Kolmogorov--Smirnov (KS) test. In each case, the KS test yields a probability $\ll$5\% that the two samples are drawn from the same distribution, confirming that the differences are statistically significant.

In the bottom panels of Figure \ref{fig:incidence}, we examine how the detection rate of ionized gas outflows varies with the physical properties of the host galaxies. There is a strong, monotonic rise in the detection rate of outflows with increasing SFR, sSFR, and \SSFR. Specifically, the detection rate climbs from $\sim$3\% for galaxies with log(SFR/M$_{\odot}\, \text{yr}^{-1}$) $<$ 0.7 to 57\% in the highest SFR bin, from $\sim$1\% at log(sSFR/yr$^{-1}$) $< -10.4$ to 27\% in the highest bin, and from $\sim$1\% at log($\Sigma_{\text{SFR}}$/yr$^{-1}$ kpc$^{-2}$) $< -1.8$ to 63\% in the highest bin. For \Mste\ and \sSSFR, the outflow detection rate first peaks at 13\% and 8\%, respectively, before dropping steeply to 5\% and 5\% in the highest bins. In contrast, the detection rate in inclination is relatively flat across a wide range, with average detection rate of 3\%.

Unsurprisingly, the detection of ionized outflows is strongly associated with elevated star-formation activity (SFR, sSFR, and \SSFR). This connection is expected, as galactic outflows from star-forming galaxies are thought to be driven by the energy and momentum injected into the ISM during the late stages of massive stellar evolution. Similar trends of outflows becoming more common at higher levels of star-formation activity have been reported for ionized outflows at low and high redshifts \citep[e.g.,][]{Avery21, Weldon24} and in cool, neutral outflows traced by absorption lines \citep[e.g.,][]{Roberts20}. It is important to note here that we detect ionized gas outflows at low \SSFR, down to log(\SSFR/$M_{\odot}\ yr^{-1}\ kpc^{-2}$) $\sim$ $-$2.4. Nearly 60\% of our outflow sample falls below the empirical threshold of log(\SSFR/$M_{\odot}\ yr^{-1}\ kpc^{-2}$) $\sim$ $-$1.3 proposed by \cite{Heckman02}, which is interpreted as the point when energy and momentum can overcome the gravity of the galaxy disk and launch an outflow. The high fraction of outflows detected in low \SSFR\ galaxies may result from the MaNGA survey's high sensitivity, which enables the detection of broad emission components even in such galaxies, or because the global \SSFR\ explored here underestimates the local \SSFR\ in small star-forming regions that launch the outflows \citep[e.g.,][]{Davies19, Lopez19, Roberts20}.

\begin{figure*}
  \includegraphics[width=\linewidth, keepaspectratio]{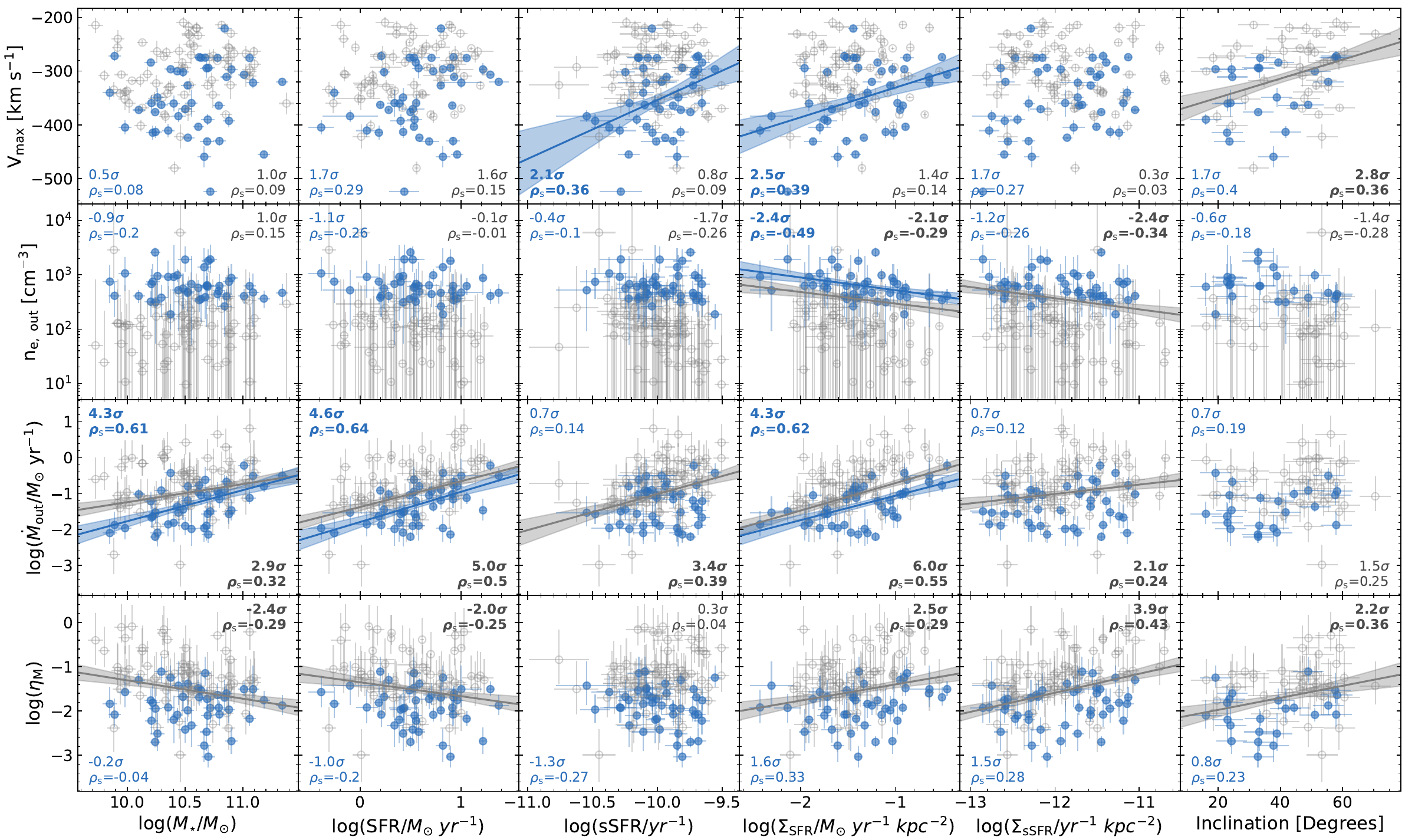}
  \vspace{-0.6cm}
  \caption{Outflow properties as a function of host galactic properties. Galaxies with outflows in the full (gold) sample sample are plotted as gray (blue) For both the full and gold samples, we list the number of standard deviations from the null hypothesis that the quantities are uncorrelated and the Spearman correlation coefficient, in gray and blue, respectively. Panels that include linear fits show where the quantities are correlated at $\geq$2$\sigma$, in gray and blue, for the full and gold samples respectively.}
  \label{fig:broad_v_tot}
\end{figure*}

The abrupt drop-off in the observed outflow detection rate at high inclinations provides useful constraints on the outflow geometry in our sample. As the ISM is highly stratified, with much higher densities in the disk plane than along the vertical direction, outflows tend to emerge perpendicular to the disk along the minor axis in a biconical structure \citep{Heckman90, Katz93}. The relatively constant outflow detection rate for galaxies with inclinations below $\sim$60$^{\circ}$ is consistent with this picture, suggesting that most outflows are oriented along the minor axis and are sufficiently wide to be visible across a range of inclinations. Under the simplifying assumptions that: (a) outflows are aligned with the minor axis and (b) opening angles are independent of inclination, our results imply half-opening angles of $\lesssim$ 60$^{\circ}$. Otherwise, the observed detection rate would begin to decline at inclinations lower than 60$^{\circ}$. This estimate is broadly consistent with studies of nearby galaxies, which typically find half-opening angles in the range of 30$^{\circ}\, -\, $60$^{\circ}$ \citep[e.g.,][]{Heckman90,Veilleux05}. The rapid decline in outflow detection rate with inclination may also be caused by observational biases. For example, the amount of material observed along the line of sight increases toward edge-on systems, leading to large column densities that can obscure or attenuate any potential emission from outflowing gas. Overall, these results support a scenario in which galactic winds are wide-angle, biconical flows, whose detectability depends on inclination and opening angle.

The apparent lack of outflows in galaxies at the extreme ends of host galaxy properties (i.e., low and high stellar mass, low star-formation activity) may reflect limitations in our fitting technique or the observations rather than the actual absence of outflows from such objects. The detectability of outflows depends on several physical properties of the outflows, such as the amount of outflowing gas, the energy injected into the gas, and the extent of the outflow. As these properties decrease, which is likely for low-mass or low-SFR objects, the emission of the broad component becomes indistinguishable from that of HII regions, such that emission lines cannot be decomposed into separate components. Even when outflow properties are strong, their signatures may be masked or suppressed by the high rotational velocities or deep potential wells of their host galaxy. Therefore, our methods presented in Section \ref{sec:methods} would miss the outflows. At the same time, the outflow properties place limitations on the sensitivity and resolution of the observations required to detect such outflows. For example, the low spectral resolution of the MaNGA survey (R $\sim$ 2000) may be a contributing factor to the lack of outflows detected in low-mass or SFR objects. \cite{Marasco23} observed nearby starburst dwarf galaxies using MUSE \citep[R $\sim$ 3000;][]{MUSE}, detecting outflows down to log(\Mste/$M_{\odot}$) $\sim$ 6 and log(SFR) $\sim -4$. Similarly, \cite{Xu22} observed local extremely metal-poor galaxies with the Magellan Echellette spectrograph \citep[R $\sim $4100;][]{Mage}, finding outflows down to log(\Mste/$M_{\odot}$) $\sim$ 5.5.

\subsection{Outflow Scaling Relations}
\label{subsec:scaling}

\subsubsection{Outflow Velocity}

\begin{figure*}
  \includegraphics[width=\linewidth, keepaspectratio]{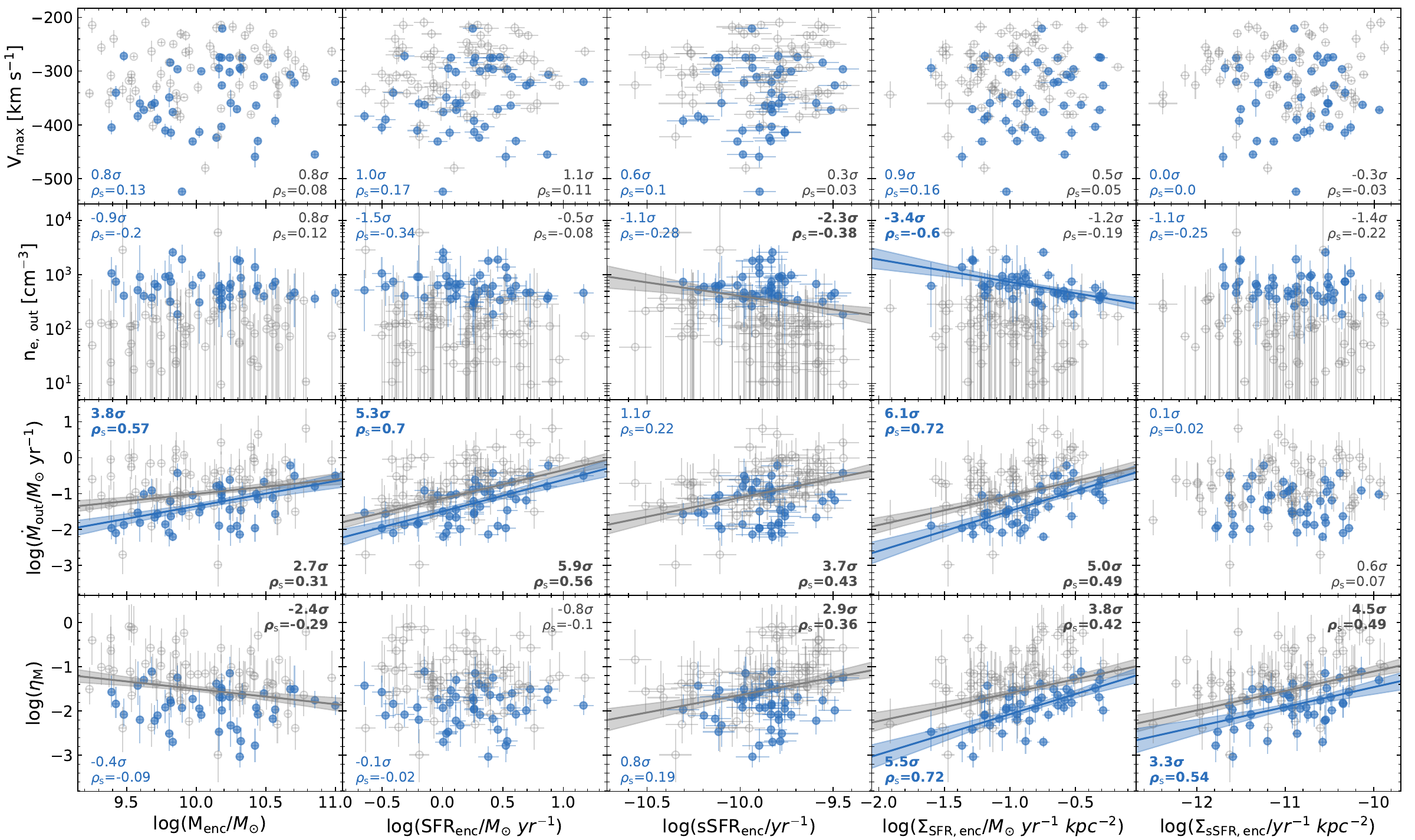}
  \vspace{-0.6cm}
  \caption{Outflow properties as a function of galactic properties enclosed within the outflow aperture. Same style as Figure \ref{fig:broad_v_tot}. As in that figure, panels that include linear fits show where the quantities are correlated at $\geq$2$\sigma$, in gray and blue, for the full and gold samples respectively.}
  \label{fig:broad_v_enc}
\end{figure*}

One of the most accessible and robustly measured properties of outflows is their velocity. As outflows from star-forming galaxies are likely driven by stellar feedback, correlations between outflow velocity and star-formation activity should naturally arise. We estimate the maximum outflow velocity from the broad component line profile following previous studies as \Vmax\ = $\Delta v_{\rm{br}}$ -- 2$\sigma_{\rm{br}}$, where $\sigma_{\rm{br}}$ is the Gaussian $\sigma$ value of the broad component \citep[see][]{Genzel11, Genzel14, Wood15}. 
The top panels of Figures \ref{fig:broad_v_tot} and \ref{fig:broad_v_enc} show the maximum outflow velocity as a function of global and enclosed galactic properties, respectively. 

Using the \texttt{linmix} package \citep{linmix}, we fit linear relations between \Vmax\ and galaxy properties for the full (gray and blue circles) and gold samples. In the full sample, we find no significant ($>$3$\sigma$) correlations with stellar mass or star formation properties. Across the range of stellar masses and star formation properties probed, outflow velocity exhibits a large $\sim$200 -- 400 km s$^{-1}$ scatter at a given fixed galactic property. This scatter seems to indicate that once the conditions for launching an outflow are satisfied, outflow velocity decouples from the properties of their host galaxy and/or launching region. Galaxies with outflows that have reliable \neout\ estimates are shown as blue circles (``gold'' sample; see Section \ref{subsubsec:ne_scaling}). With the gold sample, we find marginal correlations with sSFR (2.1$\sigma$) and \SSFR\ (2.5$\sigma$), such that faster outflows are found from galaxies with lower star formation properties.

Previous efforts to explore the correlation between outflow velocity and star formation properties have yielded conflicting results. A lack of trends between outflow velocity and star formation properties has been demonstrated in several studies \citep[e.g.,][]{Pinto19, Perrotta21, Weldon24}. Conversely, \cite{Cicone16} and \cite{Xu22} find strong, significant correlations with outflow velocity increasing with star-formation properties \citep[see also weak correlations in][]{Arribas14, Avery21}. However, such trends only begin to emerge either: (1) at high SFRs (log(SFR) $>$ 0) or (2) over large ranges in SFR ($\Delta$log(SFR) $\sim$ 4 dex). In this context, our findings are qualitatively consistent with previous results in that no strong pattern between outflow velocity and star formation properties emerges for low properties over the small dynamic range probed.

There is a marginal (2.8$\sigma$) correlation between the \Vmax\ and inclination, such that the faster outflows are found in lower-inclination (i.e., face-on) galaxies. The best-fit line, using \texttt{linmix}, is given in Table \ref{tbl:corr}. In our outflow sample, galaxies at the lowest inclinations ($i$ $<$ 23$^{\circ}$) have an average outflow velocity of $-360\pm$20 km s$^{-1}$, while those at the highest inclinations ($i$ $>$ 58$^{\circ}$) have a considerably lower average velocity of $-270\pm$20 km s$^{-1}$. This trend of faster outflows from galaxies at lower inclinations is consistent with the collimation of outflows. Within this picture, measured outflow velocities would strongly depend on inclination, with low-inclination (face-on) galaxies exhibiting faster outflows compared to high-inclination (edge-on) galaxies. Indeed, previous studies have observed faster outflows from galaxies at lower inclinations using both absorption- and emission-line tracers \citep[e.g.,][]{Chen10, Rubin14, Roberts19, Avery21}.

\subsubsection{Electron Density}
\label{subsubsec:ne_scaling}

A key parameter of outflows is their electron density, which plays a central role in estimating mass-outflow rates. In theory, the electron density of ionized gas outflows can be directly measured from the broad components of the density-sensitive \SII\ line ratio. In practice, this proves challenging, as the broad \SII\ components are often too faint to detect in individual galaxies. It remains unclear whether and how the electron density of ionized outflows depends on the physical properties of their host galaxies.

We estimate electron densities using PyNeb \citep[version 23-01;][]{Luridiana15}, adopting an electron temperature of T = 10,000 K, consistent with other ionized outflow studies. It is important to note here that this version of PyNeb incorporates updated \SII\ transition probabilities from \cite{Rynkun19} rather than the values from \cite{Podobedova09}, commonly used in previous ionized outflow studies \citep[e.g.,][]{Pinto19, Schreiber19, Swinbank19, Fluetsch21, Marasco23, Weldon24}. As a result, our electron density estimates are $\sim$44\% larger than those reported in earlier studies. Of the 115 galaxies with detected outflows, 70 (2) have inconclusive measurements as they fall within 1$\sigma$ of $\lesssim$50 cm$^{-3}$ ($\gtrsim$10$^4$ cm$^{-3}$, where the density curve begins to flatten toward the theoretical low-density (high-density) limit\footnote{We note that our main conclusions do not change if the low- and high-density limits proposed by \cite{Kewley19} are adopted.}. Here, we present both results for the 115 galaxies with detections (hereafter the ``full sample'') and for the 43 galaxies with the most reliable \neout\ estimates (hereafter the ``gold sample'').

In the second row of Figures \ref{fig:broad_v_tot} and \ref{fig:broad_v_enc}, we examine how \neout\ varies with the properties of their host galaxy. Using the full sample (gray and blue circles), we find marginal correlations with global \SSFR\ ($-2.1\sigma$) and \sSSFR\ ($-2.4\sigma$), such that denser outflowing gas is found in galaxies with lower star-formation activity. Considering only outflows in the gold sample (blue circles), we find that \neout\ is nearly flat across the range of stellar mass, star formation properties, and inclinations probed, with an interquartile range spanning from 400 to 800 cm$^{-3}$.
However, we find a marginal ($-2.4\sigma$) correlation between \neout\ and \SSFR\ in the gold sample. Moving toward smaller scales (Figure \ref{fig:broad_v_enc}), we find a marginal correlation with \sSFRenc\ ($-2.3\sigma$), while the correlation between \neout\ and \SSFRenc\ becomes significant ($-3.4\sigma$) in the gold sample.

Overall, the consistent marginal correlations found between \neout\ and star-formation-rate surface density suggest a physical connection between the properties. A decreasing \neout\ with increasing \SSFR\ differs from the reported increase in ISM electron density with \SSFR\ \citep[e.g.,][]{Shimakawa15, Reddy23, Li25}. Similarly, a recent study by \cite{Xu23a} found a significant positive correlation between global \SSFR\ and \neout\ for cool gas outflows estimated from the column density ratio of the \ion{Si}{2}$^{*}\,\lambda$1265 fine-structure excited line to the \ion{Si}{2} $\lambda$1260 absorption line. We will further discuss these results in Section \ref{subsec:ne}.

\begin{deluxetable}{cccrr}
    \tabletypesize{\scriptsize}
    \tablecolumns{5}
    \tablecaption{Scaling Relations} 
    \label{tbl:corr}
    \tablehead{\colhead{Galaxy Property} & \colhead{Outflow Property} & \colhead{Sample} & \colhead{Slope} & \colhead{Intercept}}
    \startdata
        \multicolumn{5}{c}{\text{Total Galaxy Properties}}\\
        \hline
        log(sSFR/$\rm{yr^{-1}}$)                          & \Vmax [km/s$^{-1}$]                   & Gold &    109$\pm$53    &    740$\pm$530\\
        log(\SSFR/$\rm{M_{\odot}\, yr^{-1}\, kpc^{-2}}$)  & \Vmax [km/s$^{-1}$]                   & Gold &    55$\pm$22     & $-$275$\pm$31\\
        Inclination [$^{\circ}$]                          & \Vmax [km/s$^{-1}$]                   & Full &    1.76$\pm$0.64 & $-$380$\pm$30\\
        log(\SSFR/$\rm{M_{\odot}\, yr^{-1}\, kpc^{-2}}$)  & log(\neout/$\text{cm}^{-3}$)          & Full & $-$0.21$\pm$0.11 &    2.26$\pm$0.15\\
        log(\SSFR/$\rm{M_{\odot}\, yr^{-1}\, kpc^{-2}}$)  & log(\neout/$\text{cm}^{-3}$)          & Gold & $-$0.24$\pm$0.10 &    2.48$\pm$0.14\\
        log(\sSSFR/$\rm{yr^{-1}\, kpc^{-2}}$)             & log(\neout/$\text{cm}^{-3}$)          & Full & $-$0.21$\pm$0.09 &    0.11$\pm$1.00\\
        log(\Mste/$\rm{M_{\odot}}$)                       & log(\Mdot/$\rm{M_{\odot}\ yr^{-1}}$)  & Full &    0.51$\pm$0.18 & $-$6.29$\pm$1.94\\
        log(\Mste/$\rm{M_{\odot}}$)                       & log(\Mdot/$\rm{M_{\odot}\ yr^{-1}}$)  & Gold &    0.87$\pm$0.23 & $-$10.4$\pm$2.5\\
        log(SFR/$\rm{M_{\odot}\ yr^{-1}}$)                & log(\Mdot/$\rm{M_{\odot}\ yr^{-1}}$)  & Full &    0.72$\pm$0.16 & $-$1.37$\pm$0.10\\
        log(SFR/$\rm{M_{\odot}\ yr^{-1}}$)                & log(\Mdot/$\rm{M_{\odot}\ yr^{-1}}$)  & Gold &    0.84$\pm$0.21 & $-$1.79$\pm$0.15\\
        log(sSFR/$\rm{yr^{-1}}$)                          & log(\Mdot/$\rm{M_{\odot}\ yr^{-1}}$)  & Full &    1.00$\pm$0.31 &    9.0$\pm$3.1\\
        log(\SSFR/$\rm{M_{\odot}\, yr^{-1}\, kpc^{-2}}$)  & log(\Mdot/$\rm{M_{\odot}\ yr^{-1}}$)  & Full &    0.77$\pm$0.15 &    0.06$\pm$0.20\\
        log(\SSFR/$\rm{M_{\odot}\, yr^{-1}\, kpc^{-2}}$)  & log(\Mdot/$\rm{M_{\odot}\ yr^{-1}}$)  & Gold &    0.68$\pm$0.18 & $-$0.38$\pm$0.24\\
        log(\sSSFR/$\rm{yr^{-1}\, kpc^{-2}}$)             & log(\Mdot/$\rm{M_{\odot}\ yr^{-1}}$)  & Full &    0.26$\pm$0.13 &    2.1$\pm$1.5\\
        log(\Mste/$\rm{M_{\odot}}$)                       & log(\Mload)                           & Full & $-$0.42$\pm$0.17 &    2.8$\pm$1.8\\
        log(SFR/$\rm{M_{\odot}\ yr^{-1}}$)                & log(\Mload)                           & Full & $-$0.32$\pm$0.16 & $-$1.35$\pm$0.10\\
        log(\SSFR/$\rm{M_{\odot}\, yr^{-1}\, kpc^{-2}}$)  & log(\Mload)                           & Full &    0.37$\pm$0.15 & $-$1.04$\pm$0.21\\
        log(\sSSFR/$\rm{yr^{-1}\, kpc^{-2}}$)             & log(\Mload)                           & Full &    0.43$\pm$0.12 &    3.6$\pm$1.4\\
        Inclination [$^{\circ}$]                          & log(\Mload)                           & Full &    0.01$\pm$0.01 & $-$2.25$\pm$0.28\\
        \hline
        \multicolumn{5}{c}{\text{Enclosed Galaxy Properties}}\\
        \hline
        log(\sSFRenc/$\rm{yr^{-1}}$)                         & log(\neout/$\text{cm}^{-3}$)          & Full & $-$0.50$\pm$0.24 &   -2.37$\pm$2.35\\
        log(\SSFRenc/$\rm{M_{\odot}\, yr^{-1}\, kpc^{-2}}$)  & log(\neout/$\text{cm}^{-3}$)          & Gold & $-$0.41$\pm$0.14 &    2.46$\pm$0.12\\
        log(\Menc/$\rm{M_{\odot}}$)                          & log(\Mdot/$\rm{M_{\odot}\ yr^{-1}}$)  & Full &    0.40$\pm$0.15 & $-$5.01$\pm$1.51\\
        log(\Menc/$\rm{M_{\odot}}$)                          & log(\Mdot/$\rm{M_{\odot}\ yr^{-1}}$)  & Gold &    0.70$\pm$0.20 & $-$8.35$\pm$2.02\\
        log(\SFRenc/$\rm{M_{\odot}\ yr^{-1}}$)               & log(\Mdot/$\rm{M_{\odot}\ yr^{-1}}$)  & Full &    0.79$\pm$0.15 & $-$1.14$\pm$0.07\\
        log(\SFRenc/$\rm{M_{\odot}\ yr^{-1}}$)               & log(\Mdot/$\rm{M_{\odot}\ yr^{-1}}$)  & Gold &    0.88$\pm$0.20 & $-$1.50$\pm$0.09\\
        log(\sSFRenc/$\rm{yr^{-1}}$)                         & log(\Mdot/$\rm{M_{\odot}\ yr^{-1}}$)  & Full &    1.06$\pm$0.30 &    9.5$\pm$3.0\\
        log(\SSFRenc/$\rm{M_{\odot}\, yr^{-1}\, kpc^{-2}}$)  & log(\Mdot/$\rm{M_{\odot}\ yr^{-1}}$)  & Full &    0.82$\pm$0.18 & $-$0.24$\pm$0.17\\
        log(\SSFRenc/$\rm{M_{\odot}\, yr^{-1}\, kpc^{-2}}$)  & log(\Mdot/$\rm{M_{\odot}\ yr^{-1}}$)  & Gold &    1.12$\pm$0.23 & $-$0.37$\pm$0.20\\
        log(\Menc/$\rm{M_{\odot}}$)                          & log(\Mload)                           & Full & $-$0.34$\pm$0.14 &    1.93$\pm$1.43\\
        log(\sSFRenc/$\rm{yr^{-1}}$)                         & log(\Mload)                           & Full &    0.79$\pm$0.28 &    6.2$\pm$2.8\\
        log(\SSFRenc/$\rm{M_{\odot}\, yr^{-1}\, kpc^{-2}}$)  & log(\Mload)                           & Full &    0.63$\pm$0.18 & $-$0.97$\pm$0.16\\
        log(\SSFRenc/$\rm{M_{\odot}\, yr^{-1}\, kpc^{-2}}$)  & log(\Mload)                           & Gold &    0.91$\pm$0.20 & $-$1.16$\pm$0.17\\
        log(\sSSFRenc/$\rm{yr^{-1}\, kpc^{-2}}$)             & log(\Mload)                           & Full &    0.44$\pm$0.11 &    3.3$\pm$1.2\\
        log(\sSSFRenc/$\rm{yr^{-1}\, kpc^{-2}}$)             & log(\Mload)                           & Gold &    0.45$\pm$0.15 &    3.0$\pm$1.6\\  
    \enddata
    \tablecomments{Scaling relations between ionized gas outflows properties and galaxy properties. With the exception of \Vmax, linear regressions are performed in log-log space. All linear regressions are fit using the \texttt{linmix} package \citep{linmix}. From the resulting posterior distributions, we adopt the median slope and intercept as the fit parameters and estimate their uncertainties using the 16th and 84th percentiles.}
\end{deluxetable}

\subsubsection{Mass-Outflow Rate}
\label{subsubsec:mdot_scaling}

Now with estimates of the outflow radii, velocity, and electron density, we can estimate the mass-outflow rate (\Mdot) of these ionized gas outflows. Following the assumptions that the outflowing gas is in a cone of solid angle $\Omega$, the outflow rate and velocity are constant, and that the gas in the broad component is photoionized and in case B recombination, we adopt the simple outflow model described in \cite{Genzel11} and \cite{Newman12}:
\begin{equation}
    \label{equ:mdot}
    \dot{M}_{\rm{out}} = \frac{1.36m_{H}}{\gamma_{\text{H}\alpha} n_{e,\ out}} \left( L_{\text{H}\alpha,\  \rm{Broad}}\right) \frac{V_{\rm{out}}}{R_{\rm{out}}}
\end{equation}
where $m_{H}$ is the atomic mass of hydrogen, $\gamma_{\text{H}\alpha}$($T_{e}$) = 3.56$\times$10$^{-25}$$T_{4}^{-0.91}$ erg cm$^{-3}$ s$^{-1}$ is the \HA\ emissivity at an electron temperature $T_{4}$ = 10$^{4}$K, \neout\ is the electron density of the outflow, $L_{\text{H}\alpha, \rm{Broad}}$ is the dust-corrected \HA\ luminosity of the broad component, $V_{\rm{out}}$ is the velocity of the outflow, and $R_{\rm{out}}$ is the radial extent of the outflow. We use \ROUT, \neout, and \Vmax\ as described in previous sections. The broad component \HA\ luminosity is measured individually for each galaxy, corrected for dust using the Balmer decrement within the outflow aperture, and then scaled by the broad-to-single \HA\ ratio (F$_{\text{broad}}$/F$_{\text{single}}$).

We adopt this model to facilitate comparisons with similar studies in the literature. However, varying the assumptions of the model can have noticeable effects on \Mdot. For example, the inverse dependence of $\gamma_{\text{H}\alpha}$ on $T_{e}$ creates a temperature dependence for \Mdot. \cite{Khoram25} measured direct ISM electron temperatures from [\ion{O}{3}]$\lambda$4363 for MaNGA star-forming galaxies in bins of SFR and \Mste. They find a temperature range of $\sim$4,000 $-$ 22,000K, with an average value of 11,600K. Adopting this higher temperature for the outflowing gas would increase \Mdot\ by a factor of 1.14. Similarly, the outflow extent is a large source of uncertainty in this model. Studies of nearby galaxies have found that ionized gas outflow extents can range from hundreds of parsecs \citep[e.g.,][]{Chisholm16, Cronin25} to several kpc \citep[e.g.,][]{Watts24, Lopez25, Mazzilli25, Attwater26}. In comparison, the outflow extents measured in Section \ref{subsec:cog} have an average value of 2.3 kpc. Although, with our methodology, \ROUT\ may trace the size of the region driving the outflowing gas rather than the distance the gas has traveled from the galaxy.

The third row of Figure \ref{fig:broad_v_tot} shows the mass-outflow rate as a function of stellar mass, star formation properties, and inclination. We find significant correlations between \Mdot\ and stellar mass ($2.9\sigma$ and $4.3\sigma$), SFR ($5.0\sigma$ and $4.6\sigma$), and \SSFR\ ($6.0\sigma$ and $4.3\sigma$) for both the full and gold samples. In addition, the full sample has correlations with sSFR ($3.4\sigma$) and \sSSFR\ ($2.1\sigma$). These findings indicate that the mass-loss rate of ionized outflows increases with stellar mass, SFR, and \SSFR\ of the host galaxies, as expected. Given that these outflows are likely driven by star-formation feedback, a strong correlation between \Mdot\ and star-formation activity would naturally arise. These results are in agreement with previous ionized outflow studies. Using the DR3 of the MaNGA survey, \cite{Avery21} find that \Mdot\ correlates strongest with SFR, moderately with stellar mass, and weakest with \SSFR\ \citep[see also][]{Arribas14, Couto21, Marasco23}.

\begin{figure*}
  \includegraphics[width=\linewidth, keepaspectratio]{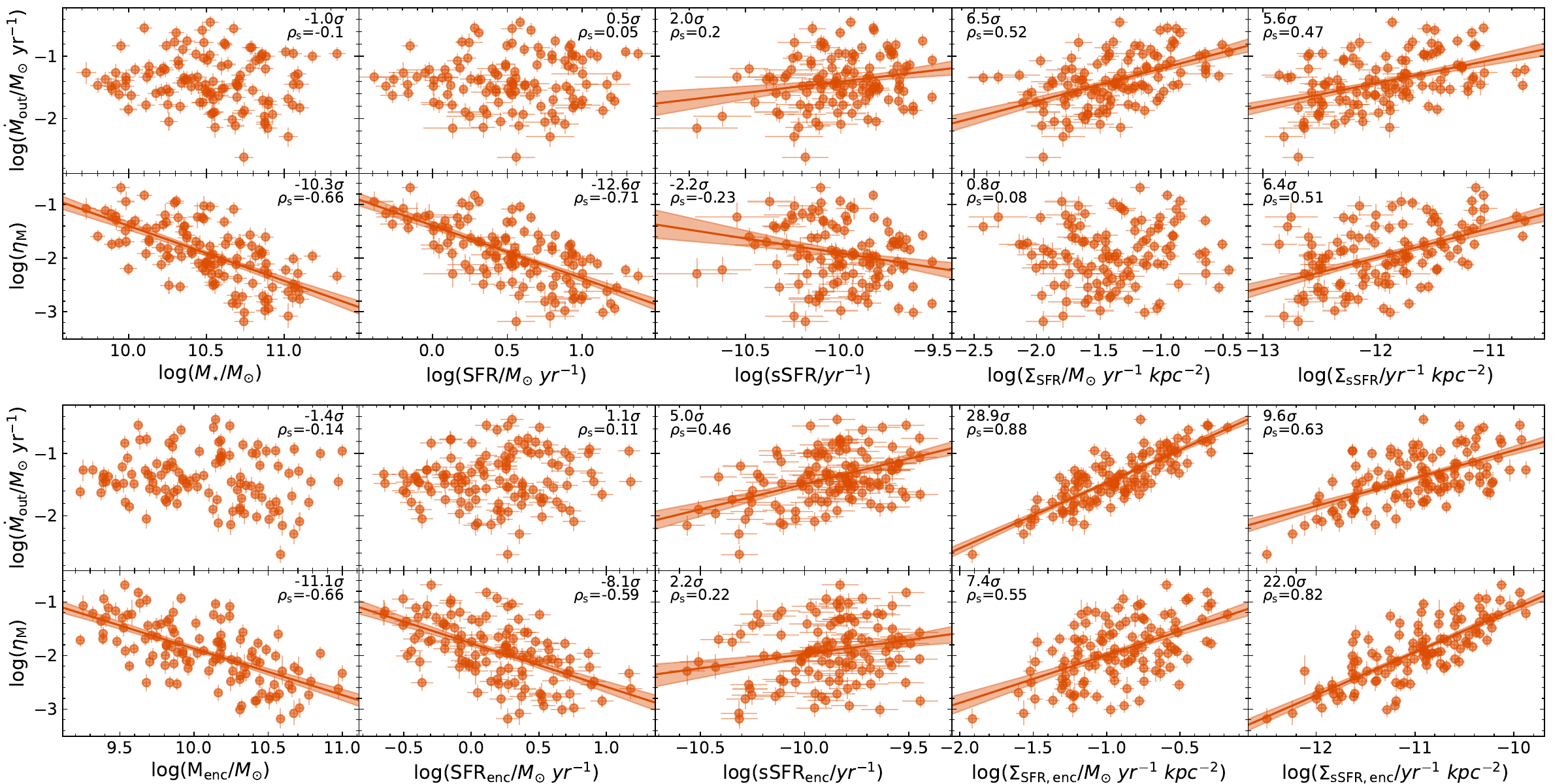}
  \vspace{-0.6cm}
  \caption{Mass-outflow rates and mass-loading factors with fixed values of \neout\ as a function of global galactic properties (top panels) and enclosed within outflow aperture (bottom panels) for the full sample. In the corner of each panel, we list the number of standard deviations from the null hypothesis that the quantities are uncorrelated and the Spearman correlation coefficient. Panels that include linear fits show where the quantities are correlated at $\geq$2$\sigma$. The scaling relations with fixed values of \neout\ are vastly different compared to with free values of \neout\ (i.e., the bottom two panels of Figures \ref{fig:broad_v_tot} and \ref{fig:broad_v_enc}).}
  \label{fig:fixed}
\end{figure*}

Moving toward smaller scales, the third row of Figure \ref{fig:broad_v_enc} shows the mass-outflow rate as a function of enclosed stellar mass and star formation properties. We again find correlations between \Mdot\ and \Menc\ ($2.7\sigma$ and $3.8\sigma$), \SFRenc\ ($5.9\sigma$ and $5.3\sigma$), and \SSFRenc\ ($5.0\sigma$ and $6.1\sigma$) for both the full and gold samples, and a correlation with \sSFRenc\ ($3.7\sigma$) in the full sample. The similar or weaker trends observed between \Mdot\ with enclosed properties, relative to the global properties, may indicate that outflows are more closely tied to the properties of their host rather than the local regions where outflows are launched. We will discuss this further in Section \ref{subsec:scale}.

\subsubsection{Mass-loading Factor}

Finally, a fundamental property of outflows is their mass-loading factor (\Mload), which represents the amount of mass carried by the outflow per stellar mass formed and, for star-formation driven outflows, is thought of as a diagnostic of outflow efficiency. Specifically, \Mload\ is defined as the mass-outflow rate normalized by the star formation rate: \Mload\ = \Mdot/SFR. Using this definition, we estimate the mass-loading factor for our sample, finding \Mload\ ranging from 0.001 -- 0.07, with a median value of 0.013. These values are on the lower range of those reported in previous studies of ionized outflows. For example, \cite{Arribas14} measured \Mload\ $\sim$ 0.1 -- 1.0 in local LIRGs, \cite{Avery21} reported \Mload\ $\sim$ 0.01 -- 1.0 using DR3 of the MaNGA survey, and \cite{Couto21} found a typical \Mload\ $\sim$ 0.12 in 13 local star-forming galaxies. The lower \Mload\ values of our sample likely reflect the higher outflow electron densities of our sample compared to those assumed in these earlier works.

The bottom row of Figure \ref{fig:broad_v_tot} displays the relations between \Mload\ and global galactic properties. With the full outflow sample, we find correlations between \Mload\ and \Mste\ ($-2.4\sigma$), SFR ($-2.0\sigma$), \SSFR\ ($2.5\sigma$), \sSSFR\ ($3.9\sigma$), and inclination ($2.2\sigma$). However, these correlations are not found when considering only galaxies in the gold sample. Instead, \Mload\ exhibits a consistent 1 dex scatter, regardless of the galactic property, which dominates over any underlying trend. The lack of correlations may, in part, be due to the narrow range of properties probed by our outflow sample, such that we would only be sensitive to very strong scaling relations. This is consistent with previous observational studies that report no correlations or only weak scaling relations between \Mload\ and galaxy properties \citep[e.g.,][]{Arribas14, Avery21, Xu22}. Similarly, cosmological simulations predict shallow scaling relations, such as \Mload\ $\propto M_{\star}^{-0.35-0.45}$ or \Mload\ $\propto \Sigma_{\text{SFR}}^{-0.1-0.5}$, which would require samples covering several orders of magnitude in host galaxy properties to robustly detect \citep[][]{Muratov15, Nelson19, Kim20, Pandya21}. 

The bottom row of Figure \ref{fig:broad_v_enc} shows \Mload\ as a function of enclosed stellar mass and star formation properties. On these smaller scales, we now find significant correlations between \Mload\ and \SSFRenc\ ($3.8\sigma$ and $5.5\sigma$) and \sSSFRenc\ ($4.5\sigma$ and $3.3\sigma$) for both the full and gold samples, and marginal correlations with \Menc\ ($-2.4\sigma$) and \sSFRenc\ ($2.9\sigma$) in the full sample. These findings indicate that the local \SSFR\ and stellar mass are key properties in regulating the efficiency of outflows, with \Mload\ increasing toward higher surface densities. However, care must be taken when interpreting these trends. The correlations between \Mload\ and the enclosed surface densities are likely due, in part, to the shared dependence between \ROUT\ and \RENC. 

\subsection{Fixed Outflow Electron Density}
\label{subsec:fixed}

Previous observational studies are often limited to a single outflow electron density estimate for entire galaxy samples, potentially introducing biases into estimates of \Mdot\ and \Mload. Here, we investigate how this simplification affects their scaling relations with global and local galaxy properties. Using Equation \ref{equ:mdot}, we re-estimate \Mdot\ and \Mload\ for the outflow sample, adopting the median outflow electron density of galaxies in the full sample (\neout\ = 240$^{+70}_{-60}$ cm$^{-3}$). Using this \neout, the resulting fixed-density estimates, \Mdotfixed\ and \Mloadfixed, are systematically lower and more uniform than \Mdot\ and \Mload, with median values (scatter) decreasing by $\sim$2.7$\times$ ($\sim$12$\times$) and $\sim$2.8$\times$ ($\sim$4.7$\times$), respectively.

The consequences of a fixed \neout\ extend beyond simple offsets. As shown in Figure \ref{fig:fixed}, the scaling relations of \Mdotfixed\ and \Mloadfixed\ with global (top panel) and local (bottom panel) properties are vastly different from those with \Mdot\ and \Mload\ (i.e., the bottom two panels of Figures \ref{fig:broad_v_tot} and \ref{fig:broad_v_enc}). Adopting a fixed \neout\ erases the previously significant ($>3\sigma$) correlations between \Mdotfixed\ with global and enclosed stellar mass and SFR. At the same time, previously marginal correlations between \Mloadfixed\ with global and enclosed properties are now highly significant. 

These correlations do not appear to be physically meaningful, but rather arise as a consequence of the definitions. Once the scatter from varying \neout\ is removed, \Mloadfixed\ becomes dominated by its dependence on SFR, scaling as \Mloadfixed\ $\propto$ SFR$^{-0.99\pm0.1}$. Similarly, the stronger correlations of \Mdotfixed\ and \Mloadfixed\r with the global and enclosed \SSFR\ and \sSSFR\ appear to result from reduced scatter and an underlying shared dependence between \Mdotfixed, \Mloadfixed, and surface densities with \RE, \ROUT, and \RENC.

These results demonstrate that adopting a single outflow electron density for an entire sample can severely distort the inferred relationships between outflow properties and their host galaxies. Fixed-density assumptions suppress trends with \Mdot\, while introducing spurious correlations with \Mload. Incorporating outflow-specific electron density estimates is therefore essential to robustly constrain \Mdot, \Mload, and their scaling relations, and place meaningful observational constraints on feedback models.

\section{Discussion} 
\label{sec:discussion}

\subsection{Outflow Electron Density}
\label{subsec:ne}

\subsubsection{Outflow Electron Density and Star-Formation-Rate Surface Density}
\label{subsubsec:ne_ssfr}

Our analysis reveals marginal correlations between the electron density of ionized gas outflows and global \SSFR\ and \SSFR\ within the outflow extent, such that lower \SSFR\ tends to drive outflows with \textit{higher} gas densities. This trend appears to be the opposite of that observed for ISM electron density, which increases with host \SSFR\ \citep[e.g.,][]{Shimakawa15, Reddy23, Topping25}. Naively, higher-\SSFR\ regions, characterized by denser ISM and stronger feedback, could be expected to produce outflows with higher densities, but this is not observed. Below, we explore possible physical mechanisms for this counterintuitive observed trend.

One plausible explanation is that the internal structure of the outflows may vary with \SSFR. Rather than being homogeneous, outflows are thought to be multiphase and stratified, with cooler, denser clumps embedded within a hotter, diffuse volume-filling medium \citep[e.g.,][]{Kim18, Fielding22}. Simulations indicate that stronger feedback, anticipated in higher-\SSFR\ regions, can enhance the shredding and mixing of cold, dense clumps into the hot, diffuse component \citep[see discussion in][]{Thompson24}. This leads to lower overall densities of the ionized outflowing gas. Regions of low \SSFR\ with weaker feedback likely retain more entrained, dense clumps within the outflow, leading to higher \neout\ estimates. In this case, different internal structures could dominate the observed broad emission line components: well-mixed, lower-density gas in high-\SSFR\ regions versus multiple, denser outflowing clumps in low-\SSFR\ regions. 

To test whether this is the case for our sample, we explored whether \neout\ decreases with outflow velocity, as broad emission dominated by well-mixed gas would likely exhibit faster outflow velocities than those dominated by denser, entrained clumps. We find no significant correlation. This could suggest that the internal structure of the outflows cannot fully explain the observed relation between \neout\ and \SSFR. On the other hand, this is a very simple test, which may be affected by other factors, such as the geometry (i.e., inclination) of the outflowing gas.

\begin{figure*}
  \includegraphics[width=\linewidth, keepaspectratio]{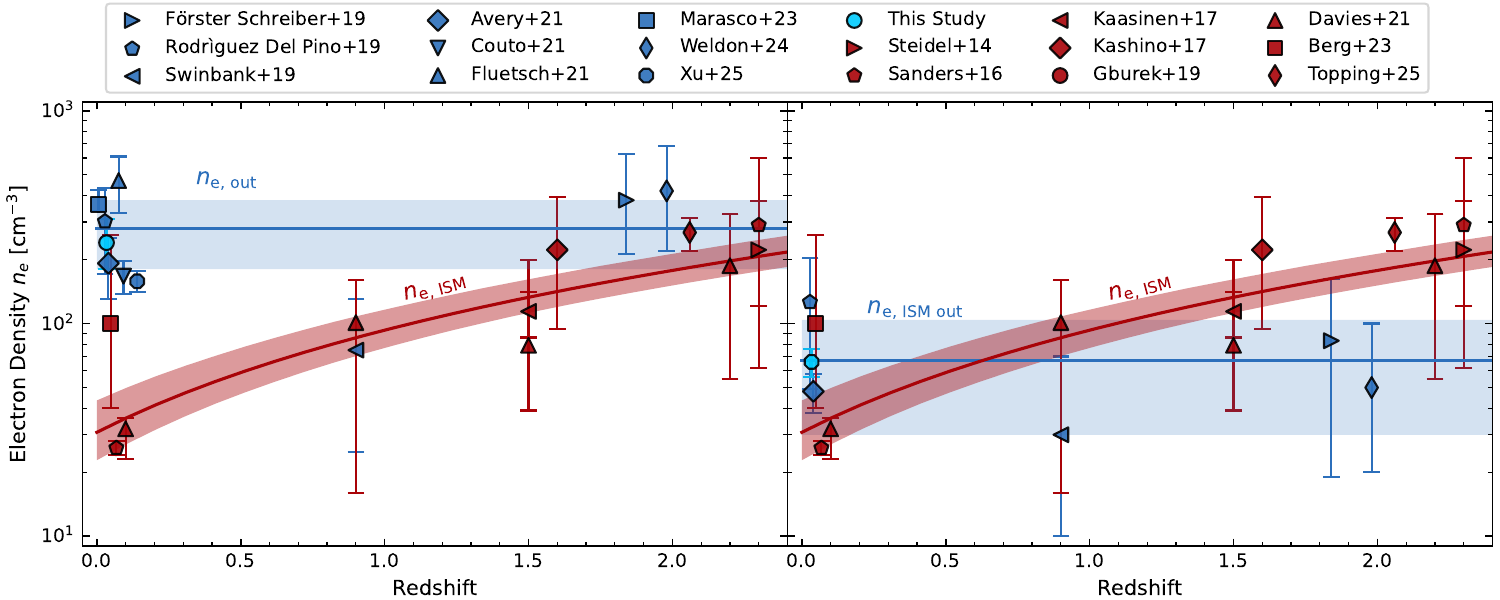}
  \vspace{-0.6cm}
  \caption{Electron density as a function of redshift. Red symbols show the median $n_{\rm{e,\, ISM}}$ of star-forming galaxies taken from the literature \citep[][]{Steidel14, Sanders16, Kaasinen17, Kashino17, Gburek19, Davies21, Berg22, Topping25}. The red line represents the best-fit power-law function in the form $n_{\text{e}}$ = (31$^{+13}_{-8}$ cm$^{-3}$)$\times$(1 + $z$)$^{1.6\pm0.2}$, with the red shaded area showing the 1$\sigma$ uncertainty. 
  \textit{Left}: Blue symbols show the median \neout\ from star-forming galaxies for our study and taken from the literature \citep[][]{Schreiber19, Pinto19, Swinbank19, Avery21, Couto21, Fluetsch21, Marasco23, Weldon24, Xu25}. The blue line represents the best-fit constant \neout = 280$\pm$100 cm$^{-3}$, with the blue shaded area showing the 1$\sigma$ uncertainty.
  \textit{Right}: Blue symbols show the median ISM density of star-forming galaxies that host outflows, $n_{\text{e,\ ISM\, out}}$, from our study and the literature \citep[][]{Schreiber19, Pinto19, Swinbank19, Avery21, Weldon24}. The blue line represents the best-fit constant \neout = 67$\pm$37 cm$^{-3}$, with the blue shaded area showing the 1$\sigma$ uncertainty.}
  \label{fig:ne_v_z}
\end{figure*}

An alternative explanation is that clumps entrained in the outflows expand. As the clumps move away from the galaxy, their density may decrease due to geometric dilution and adiabatic or pressure-driven expansion \citep[e.g.,][]{Revalski22, Xu23b, Cronin25}. Higher-\SSFR\ regions, with more efficient injection of energy and momentum into the ISM from stellar feedback, would allow outflowing clumps to reach larger radii and expand to a larger volume. As a result, \neout\ estimated from broad emission components would shift toward lower values due to the inclusion of large, lower-density clumps at large radii in the integrated emission line profiles. In contrast, lower-\SSFR\ regions would produce more compact outflowing clumps where \neout\ remains relatively high. 

To test this possibility, we explore whether \neout\ correlates with \ROUT, but we find no significant trend. This lack of correlation suggests that this simple picture of density dilution with outflow size cannot fully describe the observed inverse relation. However, as discussed in Section \ref{subsubsec:mdot_scaling}, \ROUT\ may trace the size of the region launching the outflowing gas, rather than the outflow extent.

In reality, the situation is likely complex, and many mechanisms likely operate simultaneously. The observed inverse relation between \neout\ and \SSFR\ likely reflects a complicated interplay between the structure and evolution of outflows, \SSFR, and local ISM conditions. Further constraints from spatially resolved observations and multi-phase outflow tracers are needed to disentangle these effects and clarify the physical origin of the observed trend. IFU observations at higher spatial resolution could directly trace density and velocity gradients within outflows and test whether low-density emission corresponds to large-scale expansion or different internal structures. For example, using MUSE observations, \cite{Mazzilli25} reported that in NGC 4666 \neout\ decreases out to $\sim$2 kpc from the midplane, then increases and plateaus at $\sim$3 kpc. Additionally, comparing observed electron densities to synthetically estimated \SII\ densities from cosmological simulations could help identify the physical mechanisms responsible and constrain subgrid feedback prescriptions in galaxy evolution models.

\subsubsection{Redshift Evolution}
\label{subsec:redshift}

Since the launch of JWST, several studies have shown that the electron density of the ISM increases with redshift out to $z \sim 10$ \citep[][]{Isobe23, Abdurrouf24, Topping25}. These measurements typically use optical diagnostics, such as [\ion{O}{2}] and \SII, to trace ISM electron density. This evolution is often interpreted as a natural consequence of two factors: (1) the size evolution of galaxies, as star-forming regions become denser as galaxies become more compact at higher redshifts\citep[see discussion in][]{Isobe23}, and/or (2) the metallicity evolution of galaxies, as the mean density of star-forming regions decreases with metallicity, while metallicity itself decreases with redshift \citep[see discussion in][]{Abdurrouf24}.

It is important to understand how the electron density of outflows launched from star-forming regions connects to local ISM conditions and how these evolve with redshift. If \neout\ follows a similar redshift evolution as ISM electron density, outflow efficiency would decrease toward higher redshifts (\Mload\ $\propto$ \neout$^{-1}$). If outflow properties decouple from those of their local launch regions, their electron densities could become lower than the ISM electron density, thus maintaining high efficiency at higher redshifts. Understanding how \neout\ may evolve with redshift is therefore crucial for constraining the efficiency and impact of feedback over cosmic time.

Here, we place our estimates of outflow electron density into the broader context of galaxy evolution by comparing them to estimates out to $z \sim 1.9$. We compile electron density estimates of the ISM ($n_{\rm{e,\, ISM}}$), outflows (\neout), and the ISM of galaxies hosting ionized outflows ($n_{\rm{e,\, ISM\, out}}$) from star-forming galaxies as reported in the literature and present them in Figure \ref{fig:ne_v_z}. For literature studies that report electron densities of multiple individual galaxies for a given redshift range, we adopt the median value. As shown in previous work \citep[][]{Isobe23, Abdurrouf24, Topping25}, $n_{\rm{e,\, ISM}}$ increases with increasing redshift, rising from $\sim$33 cm$^{-3}$ at $z \sim 0.03$ to $\sim$170 cm$^{-3}$ at $z \sim 1.9$. In contrast, our work here shows that neither \neout\ (left panel) or $n_{\rm{e,\, out\, ISM}}$ (right panel) exhibits any significant redshift evolution over the same range. 

Figure~\ref{fig:ne_v_z} shows that \neout\ appears relatively constant with redshift from $z \sim 0-1.9$, with a value of \neout\ = 280$\pm$100 cm$^{-3}$. Furthermore, it is elevated relative to the typical ISM density over this redshift range. As a result, the ratio between outflow density and the ISM density of the overall galaxy population decreases significantly with redshift, from $\sim$8.5$\times$ at $z \sim 0.03$ to $\sim$1.6$\times$ at $z \sim 1.9$. On the other hand, $n_{\rm{e,\, out\, ISM}}$ remains marginally (2.0$\sigma$) lower than \neout\ between $z \sim 0-1.9$, with a typical value of $n_{\rm{e,\, out\, ISM}}$ = 67$\pm$37 cm$^{-3}$. At low redshifts, $n_{\rm{e,\, out\, ISM}}$ is similar to $n_{\rm{e,\, ISM}}$ reported for the high-specific SFR galaxies of the CLASSY survey \citep{Berg22}, selected as local high-$z$ analogs. At $z \sim 1.9$, $n_{\rm{e,\, out\, ISM}}$ is less dense than the ISM density of the overall galaxy population, with a 2.4$\sigma$ difference. Although, the trends of \neout\ and $n_{\rm{e,\, out\, ISM}}$ with redshift rely on small samples of galaxies at $z >$ 0.15. Larger samples of galaxies at higher redshifts would be needed to determine whether the observed lack of evolution in the density is in fact unevolving with redshift.

The lack of an observed redshift evolution in \neout, despite the clear evolution in $n_{\rm{e,\, ISM}}$, may reflect evolution in either the outflow properties themselves or in the physical conditions under which they are launched. After emerging from the regions that launch them, outflows may undergo rapid expansion, cooling, and mixing with ambient material, causing their electron densities to quickly decouple from that of their host region. If the internal structure of outflows reaches a quasi-equilibrium state set by the balance between thermal pressure, turbulence, and radiative cooling, this could lead to a relatively uniform electron density across cosmic time. In this scenario, the constant outflow electron density could suggest that the mass-loss rate of ionized gas outflows increases with redshift, as SFR increases with redshift, enhancing their ability to suppress star formation in high-$z$ galaxies. Previous observational studies \citep[e.g.,][see also Section \ref{subsec:scale}]{Davies19, Roberts20, Chu22} have shown that outflow properties are closely tied to those of their host regions, implying that strong post-launch evolution is unlikely to be the primary factor responsible. Alternatively, stellar feedback processes may self-regulate to produce outflows with similar physical properties over cosmic time. For example, the coupling efficiency of energy injected by supernovae into surrounding gas may vary as a function of local gas density or structure \citep[e.g.,][]{Martizzi15, Hayward17}, which may lead to roughly constant outflow densities. Such regulation would challenge simple scaling relations between global ISM properties and outflow properties, indicating more complex, non-linear dependencies.

There are some caveats to these findings that can be addressed with future observations. One caveat is that our analysis is limited to outflows with sufficiently bright emission features, allowing for the decomposition and detection of separate narrow and broad emission components. If there are lower surface-brightness outflows, our analysis would miss these, as they are below current detection limits. Studying the frequency of such outflows observationally and theoretically, and estimating their gas densities will be important for future progress. A second caveat is that, as discussed in Section \ref{subsec:detection}, outflows are preferentially detected in galaxies with elevated \SSFR. This could lead to a possible bias in current samples toward compact, gas-rich systems with similar ISM conditions across redshift. At higher redshifts, this challenge is compounded by the reliance on composite spectra of galaxies with outflow signatures to estimate electron densities \citep[e.g.,][]{Schreiber19, Weldon24}, which may bias their samples toward high-density outflows. Future progress will require spatially resolved studies of larger and more representative galaxy samples, leveraging the capabilities of high-sensitivity IFUs on large telescopes.  

\subsection{Local or Global?}
\label{subsec:scale}

Whether local conditions or global galaxy properties primarily govern the properties of galactic outflows remains an open question with important implications for galaxy evolution. If local processes dominate, then regulating star formation and enriching the CGM depends sensitively on the structure and compactness of star-forming regions. If global properties set outflow properties, then feedback may act more coherently, influencing baryon cycling and galaxy quenching on larger scales. Disentangling the relative importance of local versus global drivers of outflow properties is therefore essential to fully understand how feedback shapes the evolutionary pathways of galaxies. 

\begin{figure}
  \includegraphics[width=\columnwidth, keepaspectratio]{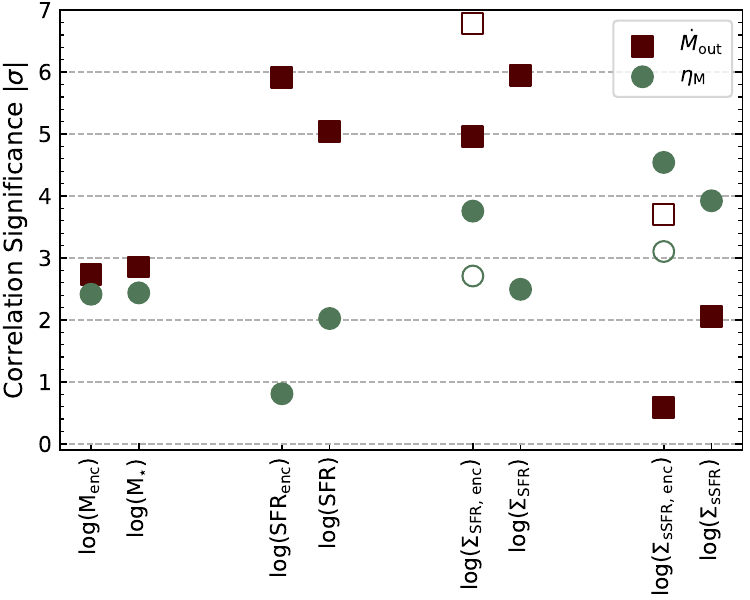}
  \vspace{-0.6cm}
  \caption{Significance level from bootstrapped pairwise Spearman correlation (closed symbols) and partial correlation (open symbols) tests using the full sample. The results for the mass-outflow rate (\Mdot) and mass-loading factor (\Mload) of the outflows are plotted as maroon squares and green circles, respectively. Both outflow properties are tested against enclosed (measured within the outflow radius) and global stellar mass, SFR, SFR surface density (\SSFR), and specific SFR surface density (\sSSFR).}
  \label{fig:local_v_global}
\end{figure}

Here we investigate whether the properties of ionized outflows in our full sample are primarily driven by enclosed or global galactic properties by performing Spearman correlation tests with bootstrapping. Figure~\ref{fig:local_v_global} shows the significance of correlations between the mass-outflow rate (\Mdot; maroon squares) and mass-loading factor (\Mload; green circles) of the outflows to galaxy properties. There appear to be more significant correlations ($>$3$\sigma$) between outflow properties and enclosed galactic properties measured within \RENC, compared to global galactic properties. At face value, this supports the conclusion that outflow properties are set by local conditions, but this interpretation does not hold across all properties. Both \Mdot\ and \Mload\ have similar correlations with global and enclosed \Mste, and \Mdot\ (\Mload) is strongly (weakly) correlated with both enclosed and global SFR. On the other hand, \Mdot\ has stronger correlations with global surface densities, while \Mload\ has stronger correlations with local surface densities. However, \Mdot, \Mload, and enclosed surface densities have shared dependencies with \ROUT\ and \RENC, raising the possibility that this shared dependence drives the observed correlations.

To test this possibility, we perform partial correlation tests with bootstrapping while controlling for \ROUT\ for these relations. We find that the correlation between \Mdot\ and \SSFRenc\ remains highly significant (6.7$\sigma$) and the correlation with \SSFRenc\ becomes significant (3.7$\sigma$), supporting a physical link between local surface density and outflow strength. In contrast, the partial correlations between the \Mload\ and \SSFRenc\ (\sSFRenc) decrease to 2.7$\sigma$ (3.1$\sigma$).

Overall, our results support a scenario in which a combination of local and global properties influence outflow properties. The strongest correlation, after controlling for \ROUT, between \Mdot\ and \SSFRenc\ is consistent with previous spatially resolved studies that identify local \SSFR\ as a key driver of outflow activity \citep[e.g.,][]{Davies19, Roberts20, Chu22}. In contrast, the similar correlations between \Mdot\ and global and enclosed \Mste, together with the stronger correlations between \Mload\ and global properties, suggest that outflow properties have a more global origin \citep[see the review by][]{Rupke18}. Simulations likewise offer mixed perspectives: high-resolution, local patch and zoom-in simulations emphasize the role of local feedback processes and ISM structure in producing realistic outflows \citep[e.g.,][]{Muratov15, Kim18, Kim20}, while large-volume cosmological simulations, which model feedback via subgrid prescriptions that couple outflows to global properties, are able to reproduce key galactic scaling relations \citep[][]{Schaye15, Nelson19, Dave19}. 

Together, our results indicate that outflow properties are sensitive to small and large scales: \Mdot\ appears more locally regulated than globally, while local and global properties influence \Mload. These results highlight the need for galaxy formation models to incorporate feedback prescriptions that account for processes operating across multiple spatial scales.

\subsection{Comparison to Simulations}
\label{subsec:simulation}

The relation between stellar mass and the mass-loading factor represents a key test of stellar feedback models, as it links small-scale driving processes to galaxy-wide regulation of star formation. In principle, these relations provide a common ground for comparing theoretical predictions to observations. In star-forming galaxies, several driving mechanisms have been proposed, including momentum injection from supernovae and stellar radiation \citep[e.g.,][]{Murray05, Murray11}, mechanical energy from supernovae \citep[e.g.,][]{Chevalier85, Strickland09}, and pressure from cosmic rays \citep[e.g.,][]{Socrates08, Uhlig12, Girichidis16}, each of which may dominate under different physical conditions. Simple analytic arguments predict that the mass-loading factor for momentum-driven outflows scales weakly with velocity and stellar mass, as \Mload\ $\propto$ $M_{\star}^{-1/3}$, while, for energy-driven outflows, the mass-loading factor is predicted to scale as \Mload\ $\propto$ $M_{\star}^{-2/3}$. Here, we place our MaNGA results in the context of both previous observations and predictions from simulations.

\begin{figure}
    \includegraphics[width=\linewidth, keepaspectratio]{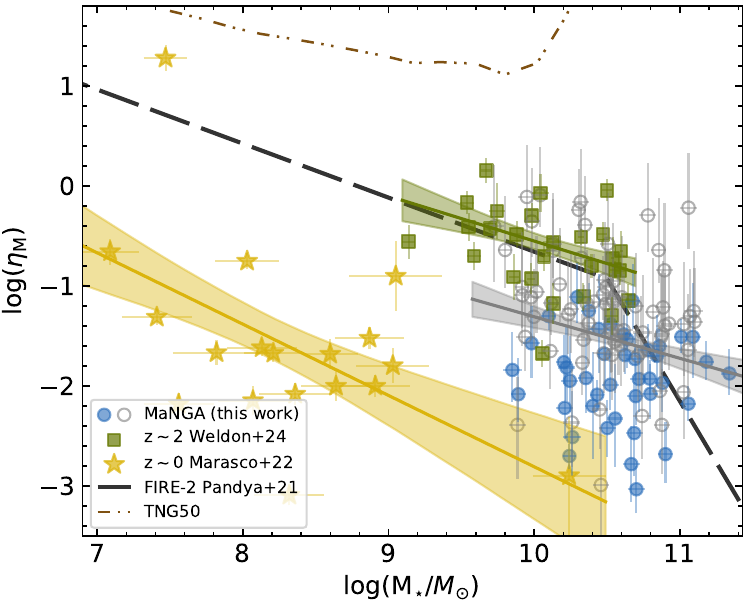}
    \vspace{-0.6cm}
    \caption{Mass-loading factor (\Mload) as a function of stellar mass. Symbols show individual MaNGA galaxies (open and blue circles), local starburst dwarf galaxies from \citet[][yellow stars]{Marasco23} and $z \sim 2$ star-forming galaxies from \citet[][green squares]{Weldon24}. Solid lines and shaded regions (68\% confidence intervals) represent the best-fit line for each population. Lines show theoretical predictions from the FIRE-2 \citet[][black dashed]{Pandya21} and Illustris TNG50 \citet[][brown dashed-dotted]{Nelson19} cosmological simulations.}
    \label{fig:eta_v_mass}
\end{figure}

Figure \ref{fig:eta_v_mass} presents \Mload\ as a function of stellar mass for galaxies in the full sample, along with comparisons to results from the literature. We find \Mload\ $\propto$ $M_{\star}^{-0.42\pm0.17}$. This scaling is intermediate between the energy- and momentum-driven cases, suggesting that these ionized outflows are driven by a combination of mechanical energy and radiation pressure. While analytical arguments and simulations predict $d\log \eta_m/d\log M_\ast < 0$, observational studies yield mixed results. Several local studies of ionized gas outflows find no significant correlation between \Mload\ and stellar mass \citep[][]{McQuinn19, Avery21, Xu22}, consistent with the lack of trend in the gold sample. Similarly, at $z \sim 1-2$, \cite{Schreiber19} and \cite{Concas22} find that \Mload\ remains roughly constant across 9 $<$ log(\Mste/$M_{\odot}$) $<$ 11 in stacks of star-forming galaxies binned by stellar mass. In contrast, other studies have identified weak, negative correlations: \cite{Arribas14} reported \Mload\ $\propto$ $M_{\rm{dyn}}^{-0.43}$ for local luminous infrared galaxies, \cite{Marasco23} found \Mload\ $\propto$ $M_{\star}^{-0.71}$ for local starburst dwarfs, and \cite{Weldon24} measured \Mload\ $\propto$ $M_{\star}^{-0.45\pm0.06}$ for $z \sim 2$ star-forming galaxies. This tension among observational studies likely stems from differences in methodology, including how outflow components are identified, assumptions about outflow geometry, and choices for \ROUT\ and \neout\ used in estimating \Mload.

Figure \ref{fig:eta_v_mass} also compares our results to predictions from cosmological simulations. We use results from the Illustris TNG50 \citep{Nelson19, Pillepich19} and Feedback in Realistic Environments \citep[FIRE;][]{Hopkins14, Hopkins18, Hopkins23}  simulations. We focus on the TNG50 values derived from z = 0.2 galaxies using outflowing gas at a fixed distance of 10 kpc from a galaxy with radial velocity $>$ 0 km s$^{-1}$ and the FIRE-2 values for warm (10$^{3}$ $<$ T $<$ 10$^{5}$ K) outflowing gas at a fixed thickness of 0.1--0.2 virial radius \citep{Pandya21}\footnote{The average virial radius of our sample is 400 kpc, estimated using the stellar-to-halo mass relation of \cite{Behroozi19}.}. We note that this is not a direct comparison, as the simulation distances are $\sim$2--8$\times$ larger than the outflow radii used in our derivation of \Mload. As the properties of the outflowing gas (e.g., density, velocity) may change with distance, \Mload\ could also vary with distance. For example, \cite{Nelson19} measured lower \Mload\ at larger distances (see their Figure 5).

Even with these caveats, our estimates of \Mload\ for ionized gas outflows are lower than those predicted by simulations. Unsurprisingly, the theoretical values from TNG50 are $\sim$4 dex larger than our estimated MaNGA mass-loading factors. In these types of large-volume simulations, small scales are not resolved and instead rely on subgrid recipes to describe stellar feedback, which then may overpredict the efficiency of stellar feedback. Additionally, TNG50 traces the total mass-loading factor across all outflow phases, whereas our values only trace the ionized phase from rest-optical emission lines. Above log(\Mste/$M_{\odot}$) $\sim$ 10, TNG50 predicts a sharp upturn in \Mload, rising from $\sim$10 to $\sim$300 between log(\Mste/$M_{\odot}$) = 10--11.3, attributed to the increasing importance of black hole-driven outflows in more massive galaxies.  

The phase separation analysis of outflows in FIRE-2 simulations \citep{Pandya21} allows for a more direct comparison with our observations. The FIRE-2 predictions for warm outflowing gas more closely align with our observed data, with an average offset of $\sim$0.3 dex ($\sim$0.6 dex) for the full (gold) sample. Notably, the prediction has a broken power-law dependence\footnote{The broken power-law relation may be due to decreasing ISM resolution in the FIRE-2 simulations toward higher stellar masses, such that \Mload\ is underestimated.} between \Mload\ and stellar mass, with a shallower ($-$0.54$\pm$0.05) slope below and steeper ($-$2.45$\pm$0.3) slope above log(\Mste/$M_{\odot}) \sim 10.5$. This appears to form an upper envelope on the mass-loading factors in the gold sample. The remaining discrepancy between our estimates and the FIRE-2 predictions likely arises from a combination of physical and observational effects. On the theoretical side, simulations may overpredict feedback coupling efficiency by neglecting processes such as turbulence and radiative cooling that dissipate injected energy and momentum. On the observational side, assumptions about the mass-outflow rate history may impact our estimates of \Mload. Additionally, while simulations tend to average over long timescales, our observations only capture instantaneous mass-loss rates, potentially missing episodes of peak outflow activity. 

Together, these factors suggest that our observed mass-loading factors may represent lower limits, while simulation predictions may reflect upper limits. Reconciling these differences will require both higher-resolution simulations capable of resolving the multiphase ISM and observational campaigns that constrain outflows across all gas phases and over larger dynamic ranges in galaxy mass.

\subsection{Fate and Impact of Ionized Outflows}
\label{subsec:escape}

\begin{figure}
  \includegraphics[width=\columnwidth, keepaspectratio]{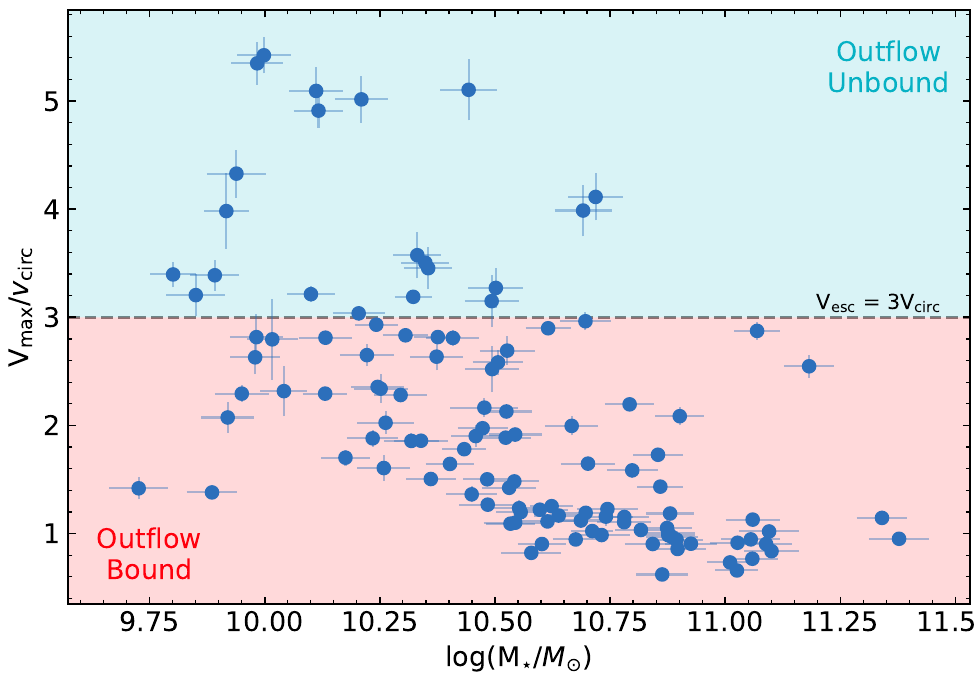}
  \vspace{-0.6cm}
  \caption{Maximum outflow velocity normalized by circular velocity as a function of stellar mass. The dashed line denotes the gas velocity required to escape the gravitational potential assuming an isothermal gravitation potential that extends to a maximum radius of $r_{\rm{max}}$, see Equation \ref{equ:escape}. Outflowing gas above 3\Vcir\ ($r_{\rm{max}}$/$r$ = 33) likely has enough velocity to escape, while below 3\Vcir\ the gas is likely retained.}
  \label{fig:escape}
\end{figure}

Large-scale galactic outflows do not impact all galaxies equally. Instead, for a given outflow power, low-mass galaxies are expected to lose more gas from outflows due to their shallower gravitational potential wells compared to higher-mass galaxies. This disparity is often cited as a key mechanism in shaping the well-known mass--metallicity relation \citep[e.g.,][]{Tremonti04}, the observed increase of gas-phase metallicity with stellar mass. In this framework, outflows in lower-mass galaxies would decrease their metallicity of their host galaxy more efficiently by directly removing metal-rich gas and indirectly by suppressing the production of new metals through star formation. In this subsection, we investigate whether the ionized outflows in our sample have sufficient speeds to escape the gravitational potential of their host or whether the gas is retained and likely recycled as a part of a galactic fountain.

For an isothermal gravitational potential truncated at $r_{\text{max}}$, the escape velocity at radius $r$ is
\begin{equation}
    v_{\text{esc}}(r) = v_{\text{circ}}\sqrt{2 \left[1 + \text{ln}\left(r_{\text{max}}/r \right) \right]}
    \label{equ:escape}
\end{equation}
where \Vcir\ is the circular velocity of the galaxy. Following \cite{Heckman15}, we estimate circular velocity as \Vcir\ = $\sqrt{2(V_{\rm{rot}}^{2} + 2\sigma^{2})}$, where $V_{\rm{rot}}$ and $\sigma$ are the observed rotation velocity and velocity dispersion of the ISM in the host galaxy, respectively. We use the \HA\ rotation velocity to estimate \Vcir\  (\texttt{HA\_GVEL\_HI\_CLIP} and \texttt{HA\_GVEL\_LO\_CLIP}; with an additional inclination correction). We use the  \HA\ velocity dispersion (\texttt{HA\_GSIGMA\_1RE}) measured as part of the MaNGA Data Analysis Pipeline.

Figure \ref{fig:escape} shows the maximum outflow velocity normalized by circular velocity as a function of stellar mass. We find a significant correlation ($-10\sigma$) between \Vmax/\Vcir\ and stellar mass, such that \Vmax/\Vcir\ decreases with increasing mass. At stellar masses below log(\Mste/$M_{\odot}$) $\sim$ 10.5, the velocity of ionized gas outflows can be several times larger than \Vcir, while above this mass outflow velocity and circular velocity are about equal. Although, there is a large amount of scatter in \Vmax/\Vcir\ for galaxies at a fixed stellar mass. To gauge whether the gas escapes, we adopt a rather conservative threshold of $r_{\text{max}}$/$r$ = 33 (\Vesc\ = 3\Vcir), such that outflowing gas likely has enough velocity to escape or is retained \citep[e.g.,][]{Veilleux05, Veilleux20}. With this definition, we find that 21 galaxies (18\%) host ionized gas outflows that can escape. These galaxies  have lower stellar masses than galaxies that retain their outflows, in agreement with other studies \citep[e.g.,][]{Schreiber19, Pinto19, Avery21}. Indeed, a KS test indicates a $p<10^{-4}$ probability that the stellar mass distributions of galaxies with and without escaping outflows are drawn from the parent distribution. 

Taken at face value, the higher escape fraction of lower-mass galaxies supports the framework that outflows can play a major role in shaping the mass-metallicity relation. However, this interpretation comes with important caveats. We lack direct measurements of the metallicity or metal-loading factor of the outflowing gas. While the metal-loading factor scales with the mass-loading factor ($\eta_{\text{Z}} \propto$ \Mload), this does not necessarily mean that outflows escaping their host galaxy are more metal-enriched than those that are retained. Although, \cite{Chisholm18} measured outflow metallicity for seven local star-forming galaxies using weak rest-frame UV absorption lines and found that the outflows from low-mass galaxies are more metal-enriched than those from higher-mass galaxies. In addition, our study only probes the warm, ionized outflow phase. Several observational studies and simulations have shown that the cold, molecular (hot) outflow phase dominates the mass (energy) outflow rates \citep[e.g.,][]{Cicone14, Kim18, Fluetsch19, Fluetsch21, Kim20, Pandya21}. Without observations of these other outflow phases, we cannot fully constrain the role that ionized gas outflows play in shaping the mass--metallicity relation.

\subsection{Impact of Shocks}
\label{subsec:shocks}

Throughout this study, we have adopted the interpretation that the broad component of rest-optical emission lines is a tracer of ionized gas entrained in star-formation driven outflows. However, the broad emission component may originate from other sources. In particular, galactic-scale outflows can drive shocks into the surrounding ISM by injecting mechanical energy that alters the ionization, temperature, and kinematic structure of the gas. As shocks propagate through the ISM, they can produce broad optical emission-line components through collisional excitation and, at high velocities, through precursor photoionization \citep[e.g.,][]{Dopita96}. The emission-line signature of a shock depends strongly on the shock velocity. In slow shocks (V $<$ 200 km s$^{-1}$), the shock front moves faster than the photoionization front, producing relatively weak high-ionization lines, but strong low-ionization lines. In contrast, fast shocks (V $>$ 200 km s$^{-1}$) create a supersonic photoionization front (“precursor”) that photoionizes gas ahead of the shock front, enhancing high-ionization lines. The emission-line ratios of shocked gas differ significantly from those of gas photoionized in H II regions, often with higher \NII/\HA\ and \OIII/\HB\ ratios similar to gas photoionized by AGN \citep[e.g.,][]{Allen08, Alarie19}.

To investigate whether shocks can explain the observed broad components, we compare our observations to MAPPINGS V \citep{Sutherland17} fully radiative shock models from the 3MdBs database \citep{Alarie19}. As all of the galaxies in our sample exhibit broad component widths $>$200 km s$^{-1}$, we consider shock+precursor models, applicable in the fast-shock regime. For each galaxy, we limit models to shock velocities within $\pm$50 km s$^{-1}$ of the observed broad-component width and to pre-shock densities less than the estimated broad component electron density. We then matched the observed broad component lines ratios -- \NII$\lambda$6585/\HA, \SII$\lambda\lambda$6718,6733/\HA, and \SII\ $\lambda$6717/$\lambda$6732 -- to the shock models, keeping the ten best-matching models per galaxy. The shock models can reproduce the observed broad component line ratios within 1$\sigma$ (3$\sigma$) for 60\% (77\%) of the sample, suggesting that shocks provide a viable explanation for the line ratios in a large fraction of the sample. To further constrain the models, we compare the observed broad component \HA\ luminosity to the model \HA\ luminosity\footnote{The shock model \HA\ luminosity is estimated by multiplying the predicted \HA\ luminosity per unit area by the area enclosed by the outflow radii.}. We find that the shock models can only reproduce the observed line ratios and broad-component \HA\ luminosity for 18\% of our sample within 1$\sigma$. This increases to 23\% of our sample if we include models within 3$\sigma$ of the measurements. For our gold sample, the shock models account for 0\% (2\%) of the sample within 1$\sigma$ (3$\sigma$) of the measurements. Therefore, while radiative shocks are a plausible origin for some of the galaxies in our sample, the majority appear inconsistent with the predictions from shock models.

In reality, the broad emission may be a mixture of outflowing gas, shocked gas, and other turbulent motions, such as turbulent mixing layers between hot and cold outflow phases. Recent studies that trace outflows using both blueshifted rest-UV absorption lines and broad components of rest-optical emission lines in individual galaxies have found that the two tracers are kinematically similar, suggesting that emission from outflowing gas dominates broad components \citep[e.g.,][]{Perrotta21, Avery22}. 

\section{Conclusions}
\label{sec:con}

We have presented an analysis of the properties of ionized gas outflows within a sample of 3555 local ($z \sim 0.04$) star-forming galaxies. Using spatially resolved spectroscopy from the final data release of the MaNGA survey, we extracted spectra within increasing aperture sizes, decomposing strong nebular emission lines into narrow and broad Gaussian components tracing virial motions within the galaxy and outflowing gas. In total, we find significant evidence for an additional broad \HA, \NII$\lambda\lambda$6550,6584 and [\ion{S}{2}$]\lambda\lambda$6718,6733 emission line component in 115 galaxies. The electron density of the outflowing gas is estimated from the broad component \ion{S}{2} line ratio, with a median value of \neout\ = 240$^{+70}_{-60}$ cm$^{-3}$. Using our estimates of outflow extent, velocity, and density in the outflow model described in \cite{Genzel11}, we estimate the mass-loss rate and mass-loading factor of the ionized gas outflows. Our main conclusions are as follows:

\begin{itemize}[leftmargin = 1em]
    \setlength\itemsep{0em}
    \item There is significant evidence for broad emission components in 3\% of the star-forming MaNGA galaxies, with their incidence increasing with the stellar mass, SFR, sSFR, and \SSFR\ of the host galaxy (Section \ref{subsec:detection}).
    
    \item We find several marginal and significant correlations between \Mdot\ of the ionized outflows and both the global properties of their host galaxies and the local properties within the outflow radius (Table \ref{tbl:corr}). In addition, for the first time, we report a marginal (significant) anti-correlation between \neout\ and global (enclosed) \SSFR, such that lower-\SSFR\ regions drive denser ionized gas outflows (Sections \ref{subsec:scaling} and \ref{subsubsec:ne_ssfr}).

    \item When \Mload\ is estimated using \neout\ on an individual basis, there are no significant correlations between \Mload\ and global properties or local properties. Adopting a single \neout\ value (i.e., the median of the sample), \Mload\ becomes significantly correlated with global and local stellar mass, SFR, \SSFR, and \sSSFR\ (Section \ref{subsec:fixed}). This suggests that outflow-specific electron density estimates are essential to robustly constrain \Mload\ and their scaling relations.

    \item Combining our outflow electron density estimates with those in literature, we find that the density of outflowing gas in the warm, ionized phase, \neout\ $\approx$280$\pm$100 cm$^{-3}$ is constant over the redshift range $z \sim 0-1.9$. This stands in sharp contrast to the strong evolution in ISM electron density of the overall galaxy population, from $\sim$33 cm$^{-3}$ at $z \sim 0$ to $\sim$170 cm$^{-3}$ at $z \sim 1.9$ (Section \ref{subsec:redshift}). The apparent invariance of outflow density across cosmic time may reflect rapid post-launch evolution or self-regulated feedback processes that produce similar densities despite evolving ISM properties.

    \item The similar significance of scaling relations between outflow properties with global and ``local'' enclosed galactic properties suggests that outflow properties are sensitive to both the properties of their host and the regions that launch outflows (Section \ref{subsec:scale}).

    \item The mass-loading factor is marginally correlated (-2.4$\sigma$) with stellar mass, scaling as \Mload\ $\propto M_{\star}^{-0.42\pm0.17}$. This scaling is intermediate between the predicted dependence for energy- or momentum-driven outflows, suggesting that these ionized outflows are driven by a combination of these mechanisms (Section \ref{subsec:simulation}).

    \item Across the sample, $\sim$18\% of the ionized gas outflows have sufficient maximum outflow velocities to escape from their host galaxy. The number of galaxies with escaping outflows increases toward lower stellar masses. At high stellar masses (log(\Mste/$M_{\odot}$) $\gtrsim$ 10.5), the outflows are retained and likely create a galactic fountain (Section \ref{subsec:escape}).

\end{itemize}

Obtaining robust constraints on the properties of outflows beyond their kinematics is crucial to understanding their impact on galaxy evolution. Here, we have leveraged IFU observations of local star-forming galaxies to study how the electron density, mass-loss rate, and mass-loading factor of ionized gas outflows vary with enclosed and global galactic properties. Our findings reinforce the idea that the physical properties of ionized outflows are influenced by the local star-forming environments that launch them. In particular, the inverse relation between \neout\ and enclosed \SSFR\ suggests that outflow velocity, entrainment, and internal structure may depend sensitively on local ISM conditions. Future progress constraining feedback will require both observational and theoretical advances. High spatial resolution IFU observations (e.g., JWST/NIRSpec, VLT/MUSE, and future 30-m class telescopes) will be essential to resolve density and velocity gradients within individual outflows and probe their internal structure. Complementary progress on the theoretical side will require cosmological and high-resolution zoom-in simulations that track the multiphase structure of outflows and provide synthetic emission-line diagnostics for direct comparison with observations. Together, these efforts will help build a better understanding of how outflow properties relate to local ISM conditions.

\begin{acknowledgments}

We would like to thank the anonymous referee for their constructive feedback that improved the paper. Funding for the Sloan Digital Sky Survey IV has been provided by the  Alfred P. Sloan Foundation, the U.S. Department of Energy Office of  Science, and the Participating Institutions.  SDSS acknowledges support  and resources from the Center for High-Performance Computing at the  University of Utah. The SDSS web site is \url{www.sdss4.org}.

SDSS is managed by the Astrophysical Research Consortium for the Participating Institutions of the SDSS Collaboration including the Brazilian Participation Group, the Carnegie Institution for Science, Carnegie Mellon University, Center for Astrophysics | Harvard \& Smithsonian (CfA), the Chilean Participation Group, the French Participation Group, Instituto de Astrofísica de Canarias, The Johns Hopkins University, Kavli Institute for the Physics and Mathematics of the Universe (IPMU) / University of Tokyo, the Korean Participation Group, Lawrence Berkeley National Laboratory, Leibniz Institut für Astrophysik Potsdam (AIP), Max-Planck-Institut für Astronomie (MPIA Heidelberg), Max-Planck-Institut für Astrophysik (MPA Garching), Max-Planck-Institut für Extraterrestrische Physik (MPE), National Astronomical Observatories of China, New Mexico State University, New York University, University of Notre Dame, Observatório Nacional / MCTI, The Ohio State University, Pennsylvania State University, Shanghai Astronomical Observatory, United Kingdom Participation Group, Universidad Nacional Autónoma de México, University of Arizona, University of Colorado Boulder, University of Oxford, University of Portsmouth, University of Utah, University of Virginia, University of Washington, University of Wisconsin, Vanderbilt University, and Yale University.

Portions of this research were conducted with the advanced computing resources provided by Texas A\&M High Performance Research Computing.

\end{acknowledgments}

\begin{contribution}


A.W. lead the analysis and writing of the paper. C.P., J.S., and R.C contributed to the interpretation.


\end{contribution}

%

\software{\texttt{astropy} \citep{Astropy13, Astropy18, Astropy22}, \texttt{pPXF} \citep{Cappellari17, Cappellari23}, \texttt{lmfit} \citep{lmfit}, \texttt{emcee} \citep{emcee}, \texttt{linmix} \citep{linmix}}

\appendix

\section{Tables}

Here we provide tables for the global galactic properties (Table \ref{tbl:global}), enclosed galactic properties (Table \ref{tbl:local}), and outflow properties (Table \ref{tbl:outflow}) for the 115 galaxies in the outflow sample.

\begin{deluxetable}{ccccccc}
    \tabletypesize{\scriptsize}
    \tablecolumns{7}
    \tablewidth{0pc}
    \tablecaption{Globally Measured Galactic Properties} 
    \label{tbl:global}
    \tablehead{\colhead{MaNGA Plate-IFU}  & \colhead{$z$} & \colhead{log($M_{\star}/M_{\odot}$)} & \colhead{log(SFR/$M_{\odot}$ yr$^{-1}$)} & \colhead{log(sSFR/yr$^{-1}$)} &
    \colhead{log$(\rm{\Sigma_{SFR}/M_{\odot}\ yr^{-1}\ kpc^{-2}})$} & \colhead{log$(\rm{\Sigma_{sSFR}/yr^{-1}\ kpc^{-2}})$}}
    \startdata
        10001-3702 & 0.0256 & 10.02$\pm$0.05 & -0.27$\pm$0.08 & -10.28$\pm$0.08 & -1.75$\pm$0.1 & -11.76$\pm$0.09\\
        10215-6102 & 0.0214 & 10.13$\pm$0.05 & 0.49$\pm$0.02 & -9.64$\pm$0.02 & -1.12$\pm$0.05 & -11.25$\pm$0.05\\
        10221-6102 & 0.0996 & 11.1$\pm$0.06 & 0.94$\pm$0.12 & -10.16$\pm$0.11 & -1.59$\pm$0.13 & -12.69$\pm$0.13\\
        \ldots     & \ldots & \ldots        & \ldots        & \ldots          & \ldots         & \ldots \\
    \enddata
    \tablecomments{Table 2 is published in its entirety in the machine-readable format. A portion is shown here for guidance regarding its form and content.}
\end{deluxetable}

\begin{deluxetable}{cccccc}
    \tabletypesize{\scriptsize}
    \tablecolumns{6}
    \tablewidth{0pc}
    \tablecaption{Locally Measured Galactic Properties} 
    \label{tbl:local}
    \tablehead{\colhead{MaNGA Plate-IFU} & \colhead{log($M_{\text{enc}}/M_{\odot}$)} & \colhead{log(SFR$_{\text{enc}}$/$M_{\odot}$ yr$^{-1}$)} & \colhead{log(sSFR$_{\text{enc}}$/yr$^{-1}$)} &
    \colhead{log$(\rm{\Sigma_{SFR,\ enc}/M_{\odot}\ yr^{-1}\ kpc^{-2}})$} & \colhead{log$(\rm{\Sigma_{sSFR,\ enc}/yr^{-1}\ kpc^{-2}})$}}
    \startdata
        10001-3702 & 9.37$\pm$0.02 & -0.48$\pm$0.08 & -9.85$\pm$0.05 & -0.87$\pm$0.09 & -10.24$\pm$0.05\\
        10215-6102 & 9.75$\pm$0.02 & 0.2$\pm$0.08 & -9.55$\pm$0.05 & -0.83$\pm$0.09 & -10.58$\pm$0.05\\
        10221-6102 & 10.51$\pm$0.02 & 0.37$\pm$0.09 & -10.14$\pm$0.05 & -1.23$\pm$0.09 & -11.74$\pm$0.06\\
        \ldots     & \ldots         & \ldots        & \ldots          & \ldots         & \ldots \\
    \enddata
    \tablecomments{Table 3 is published in its entirety in the machine-readable format. A portion is shown here for guidance regarding its form and content.}
\end{deluxetable}

\begin{deluxetable}{ccccc}
    \tabletypesize{\scriptsize}
    \tablecolumns{5}
    \tablewidth{0pc}
    \tablecaption{Measured Outflow Properties} 
    \label{tbl:outflow}
    \tablehead{\colhead{MaNGA Plate-IFU} & \colhead{$V_{\text{max}}$ [km s$^{-1}$]} & \colhead{$n_{\rm{e,\, out}}$ [cm$^{-3}$]} & \colhead{log($\dot{M}_{\rm{out}}/M_{\odot}$ yr$^{-1}$)} & \colhead{log($\eta_{m}$)} }
    \startdata
        10001-3702 & -360$\pm$18 & 120$_{-270}^{+410}$ & -1.32$_{-0.4}^{+0.56}$ & -1.06$_{-0.41}^{+0.56}$\\
        10215-6102 & -247$\pm$5 & 20$_{-90}^{+100}$ & -0.16$_{-0.35}^{+0.52}$ & -0.65$_{-0.35}^{+0.51}$\\
        10221-6102 & -287$\pm$9 & 160$_{-250}^{+380}$ & -0.42$_{-0.38}^{+0.53}$ & -1.36$_{-0.4}^{+0.54}$\\
        \ldots     & \ldots     & \ldots              & \ldots                  & \ldots \\
    \enddata
    \tablecomments{Table 4 is published in its entirety in the machine-readable format. A portion is shown here for guidance regarding its form and content.}
\end{deluxetable}



\bibliography{ref}{}
\bibliographystyle{aasjournalv7}



\end{document}